\RequirePackage{etex}

\documentclass{aa}  

\usepackage{graphicx}
\usepackage{txfonts}
\usepackage{lipsum}
\usepackage{subcaption}         
\usepackage{ulem}               
\usepackage{lscape}             
\usepackage{placeins}           
                                
\usepackage{natbib}
\bibpunct{(}{)}{;}{a}{}{,} 

\usepackage{hyperref}
\usepackage{booktabs}
\usepackage{tablefootnote}

\usepackage{multirow}

\usepackage{xcolor}

\newcommand{\new}[1]{#1}
\newcommand{\renew}[1]{#1}

\makeatletter
\let\linenumbers\relax

\let\switchlinenumbers\relax
\let\runninglinenumbers\relax
\makeatother
\begin{document}

   \title{A Kinematic History of Stellar Encounters with Beta Pictoris}
\titlerunning{Stellar Encounters with Beta Pictoris}

    \author{J.L. Gragera-Más\inst{1}\fnmsep\inst{2}
        \and S. Torres \inst{3}
        \and A.J. Mustill\inst{4}
        \and E. Villaver\inst{5}\fnmsep\inst{6}
        }

   \institute{Centro de Astrobiolog\'ia (CAB), CSIC-INTA, Camino Bajo del Castillo s/n, 28692, Villanueva de la Ca\~nada, Madrid, Spain\\
             \email{jlgragera@cab.inta-csic.es}
            \and Departamento de F\'{i}sica de la Tierra y Astrof\'{i}sica, Facultad de Ciencias F\'{i}sicas, Universidad Complutense de Madrid, E-28040, Madrid, Spain
            \and  Institute of Science and Technology Austria (ISTA), Am Campus 1, A-3400 Klosterneuburg, Austria
            \and Lund Observatory, Division of Astrophysics, Department of Physics, Lund University, Box 118, SE-221 00 Lund, Sweden
            \and Instituto de Astrof\'isica de Canarias, 38200 La Laguna, Tenerife, Spain
            \and Universidad de La Laguna (ULL), Astrophysics Department,
 38206 La Laguna, Tenerife, Spain\\ }

   \date{Received June 10, 2025}

 
  \abstract
   {Beta Pictoris is an A-type star hosting a complex planetary system with two massive gas giants and a prominent debris disk. Variable absorption lines in its stellar spectrum have been interpreted as signatures of exocomets\textemdash comet-like bodies transiting the star. Stellar flybys can gravitationally perturb objects in the outer comet reservoir, altering their orbits and potentially injecting them into the inner system, thereby triggering exocomet showers.}
   {We aim to assess the contribution of stellar flybys to the observed exocomet activity by reconstructing the stellar encounter history of $\beta$ Pictoris in the past and future.}
  {We used \textit{Gaia} DR3 data, supplemented with radial velocities from complementary spectroscopic surveys, to compile a catalogue of stars currently within 80 pc of $\beta$ Pictoris. Their orbits were integrated backward and forward in time in an axisymmetric Galactic potential (\textit{Gala} package) to identify encounters within 2 pc of the system.}
  {We identified 99\,416 stars within 80 pc of $\beta$ Pictoris at present with resolved kinematics. Among these, 49 stars (including the eight components of five binaries) encounter $\beta$ Pictoris within 2 pc between -1.5 Myr and +2 Myr. For four of the binaries, the centre-of-mass trajectories also pass within 2 pc. We estimate the sample to be more than 60\% complete within 0.5 Myr of the present.}
  {Despite $\beta$ Pictoris being the eponym of its famous moving group, none of the identified encounters involved its moving group members; all are unrelated field stars. We find no encounter capable of shaping observed disc structures, although stellar flybys may contribute to the long-term evolution of a potential Oort Cloud. Our catalogue constitutes the most complete reconstruction of the $\beta$ Pictoris encounter history to date and provides a robust foundation for future dynamical simulations.}


   \maketitle

\section{Introduction}

The $\beta$ Pictoris system (hereafter $\beta$ Pic) is one of the most iconic examples of a young, dynamically evolving planetary system, offering a valuable laboratory for investigating planet formation, disc evolution, and exocometary activity. Located at 19.6 pc from the Sun \citep{GaiaMission,GaiaEDR3Astrometry}, $\beta$ Pic is a $\sim$20 Myr-old A6V-type star \citep{Kages_Couture_2023,Gray_2006} and the namesake of the $\beta$ Pic Moving Group ($\beta\,$PMG), whose members share common kinematics and likely originated in the same star-forming region \citep{Zuckerman_2001,Kages_Couture_2023,BPMG2024aLeeR,BPMG2024bLeeJ}.

$\beta$ Pic hosts a prominent edge-on debris disc, first imaged by \cite{BPdiskdiscover}, which has been the focus of extensive study due to its structure. High-resolution observations have revealed a complex morphology, including a primary disc and an inclined secondary component \citep{Twodisks}, along with warps, clumps, and large-scale asymmetries \citep{golimowski2006hubble,lagrange2012position}. Using ALMA imaging \cite{dent2014molecular} and \cite{matra2019kuiper} traced a planetesimal belt extending from $\sim$50 to 150 au, with ongoing dust production attributed to collisional cascades and comet-like activity \citep{FEBs,GasinDiskRemains}. In addition to dust, the disc contains a significant amount of CO and atomic carbon gas \citep{gasinBP,cataldi2018alma}, which is atypical for systems of comparable age.

Recent JWST observations \citep{rebollido2024jwst} have revealed a large, asymmetric dust feature\textemdash nicknamed the ``Cat’s tail''\textemdash originating from the secondary disc. This structure likely results from a recent collisional or outgassing event and is associated with a local enhancement of circumstellar gas \citep{dent2014molecular}. Mid-infrared spectroscopy shows that the hot crystalline dust features present in 2004–2005 have since disappeared, consistent with a major collision followed by radiation-pressure blowout of small grains \citep{chen2024dustremoved}. Spectroscopic monitoring of the system has also revealed transient absorption features attributed to exocomets transiting the stellar disc \cite{ferlet1987beta,beust1990beta,kiefer2014two,lecavelier2022exocomets,ExocometsHoeijmakers2025arXiv}, reinforcing the interpretation of a highly active and dynamically rich environment.

The system also hosts two directly imaged giant exoplanets. $\beta$ Pic b, with a mass of $12\;M_\textrm{Jup}$ and an orbital distance of 10 au, was one of the first exoplanets discovered via direct imaging \citep{lagrange2009probable,lagrange2009constraining,lagrange2010giant}. $\beta$ Pic c, a closer-in giant of $9\;M_\textrm{Jup}$ at 2.7 au, was detected through a combination of radial velocity measurements and direct imaging \citep{lagrange2019evidence,nowak2020direct}. Both planets have inclinations of $\sim$89º \cite{lacour2021mass} and are aligned with the main disc but misaligned with the secondary disc \citep{rebollido2024jwst}. They were proposed\textemdash prior to their discovery\textemdash as dynamical sculptors of the observed warps and resonant structures of the surrounding disc \citep{Mouillet1997InclinedPlanet}. Numerical simulations suggest that their gravitational influence can perturb planetesimals onto star-grazing orbits via mean-motion resonances, potentially explaining the observed exocometary activity \citep{FEBs,beust2024dynamics}.

Given the presence of giant planets and an extended planetesimal disc, it is plausible that the $\beta$ Pic system hosts a distant, dynamically evolving reservoir of icy bodies\textemdash an analogue to the Solar System’s Oort Cloud. In the Solar System, the Oort Cloud is thought to have formed through a combination of scattering by giant planets and external perturbations from the Galactic tidal field and stellar encounters \citep{oort1950structure,dones2004oort,brasser2008embedded,brasser2013oort,nesvorny2018dynamical,shannon2019oort}. Several studies have used N-body simulations to explore the formation and evolution of such reservoirs \citep{levison2004sculpting,fouchard2006long,kaib2008formation,morbidelli2005origin,torres2019,vokrouhlicky2019origin,zwart2021oort}.  Unlike the Oort Cloud\textemdash which has had billions of years to reach a dynamically relaxed state\textemdash the cometary cloud around $\beta$ Pic would at present be dynamically unevolved, potentially resembling a proto-Oort Cloud in its formative stages. Studies of the influence of stellar encounters on these distant reservoirs in the Solar System \citep{RickmanCloseEncounters,garcia1999stellar,bailer2015close,bailer2018completeness,bailer2018new,torres2019} often employ the impulse approximation \citep{oort1950structure,RickmanCloseEncounters,binney2011galactic} as an alternative to full N-body simulations to study the dynamical effect induced by a fast-moving stellar perturber, reducing the three-body Sun–star–comet problem to the net impulse transferred by the star to the Sun and the comet.

\cite{kalas2001stellar} conducted an early investigation of the encounter history of $\beta$ Pic using astrometric data from the \textit{Hipparcos} catalogue \citep{perryman1997hipparcos} and radial velocities (hereafter RVs) from \cite{barbier2000catalogue}. They propagated the rectilinear trajectories of $21\,497$ stars and identified 18 candidates that passed within 5 pc of $\beta$ Pic over the past 1 Myr. Applying the impulse approximation, they assessed the effects of these encounters on the eccentricities of a hypothetical cometary cloud and concluded that, if such a reservoir exists, past stellar encounters may have contributed to the formation and shaping of an Oort Cloud-like structure.

The data releases of the European Space Agency's \textit{Gaia} mission \citep{GaiaMission,GaiaDR1,GaiaDR2, GaiaEDR3,GaiaDR3}, have enabled a major advance in our understanding of stellar dynamics in the Solar neighbourhood. With astrometry data for 1.46 billion stars and RVs for 33 million, it is now possible to reassess the encounter history of $\beta$ Pic with greatly improved precision and completeness.

The aim of this work is to construct the most complete and accurate catalogue to date of stellar encounters with the $\beta$ Pic System, using \textit{Gaia} astrometry and complementary RV data.

Section \ref{sec:catalogueSources} describes our catalogue of stars within 80 pc of $\beta$ Pic and the selection criteria applied. Section \ref{sec:CatalogueEncounters} details the orbital integration of these stars to identify past and future encounters with $\beta$ Pic; we also assess the completeness of our results using encounter probability estimates and identify binary and higher multiplicity systems within our sample. Section \ref{sec:kalas} compares our findings with the earlier study by \cite{kalas2001stellar}. Finally, Section \ref{sec:concl} presents our summary and conclusions.

\section{The \texorpdfstring{$\beta$}{beta} Pictoris neighbourhood catalogue}
\label{sec:catalogueSources}

To construct a catalogue of stars in the vicinity of $\beta$ Pic, we began with \textit{Gaia} data. The latest release, \textit{Gaia} Data Release 3 \citep{GaiaDR3} was divided into two parts: the Early Data Release 3 (hereafter GEDR3, \citealt{GaiaEDR3}), which provides astrometry \citep{GaiaEDR3Astrometry} and integrated photometry, and the full \textit{Gaia} Data Release 3 (hereafter GDR3), which superseded GEDR3 and includes additional products such as RV data \citep{GaiaDR3_RV}, BP/RP/RVS spectra, astrophysical parameter tables, and results for non-single stars, among others. 

GEDR3 also includes the \textit{Gaia} Catalogue of Nearby Stars (hereafter GCNS, \citealt{GCNS}), a census of $331\,312$ \textit{Gaia} sources that is volume-complete for spectral types earlier than M8 within a 100 pc radius centred on the Sun.

We applied the parallax correction from \cite{GaiaPLXcorr}, the proper motion correction proposed by \cite{GaiaPMcorr}, and the uncertainty corrections for both parallax and proper motion as outlined by \cite{GaiaErrorscorr}, \new{using a Python function provided by M. Pantaleoni González (priv. comm.)} to account for known systematics in the astrometric data. \new{The latter method inflates the uncertainties according to the Renormalised Unit Weight Error (RUWE) of the source, so the objects with poor astrometric solutions are assigned larger uncertainties.}

As $\beta$ Pic lies approximately $20\,\textrm{pc}$ from the Sun, the largest sphere centred on $\beta$ Pic that fits within the GCNS coverage has a radius of about $80\,\textrm{pc}$. We used the dedicated functions of the Tool for OPerations on Catalogues And Tables (hereafter TOPCAT, \citealt{TOPCAT}) to select sources within this region ({see Appendix \ref{ap:topcat}}). Our initial sample comprises $156\,995$ stars from the GCNS catalogue. Among these, $46\,321$ have complete 6D phase space information (right ascension, declination, parallax, proper motion, and RV) with associated errors and astrometric covariances. The RV data in GCNS is sourced from GDR2 and supplemented by references compiled in SIMBAD.

To incorporate the latest RV data from \textit{Gaia} \citep{GaiaDR3_RV}, we crossmatched the GCNS sources with GDR3, which is straightforward as both catalogues share the same \textit{Gaia} source ID. This increased the number of sources with RVs to $95\;453$. 

To further expand RV coverage, we crossmatched the GCNS sources with major spectroscopic surveys: The Sixth Data Release of the Radial Velocity Experiment (\textit{RAVE}, \citep{RAVEDR6_Steinmetz_2020}), the Fourth Data Release of the GALactic Archaeology with HERMES (\textit{GALAH}, \citep{GALAHDR4}), the Seventeenth Data Release of the Apache Point Observatory Galactic Evolution Experiment (\textit{APOGEE}, \citep{APOGEE_DR17}), and the Low Resolution Survey and Medium Resolution Survey of the Tenth Data Release of the Large Sky Area Multi-Object Fiber Spectroscopic Telescope (\textit{LAMOST-LRS}, \citep{LAMOST_LRS}; \textit{LAMOST-MRS}, \citep{LAMOST_MRS}). These catalogues include \textit{Gaia} IDs, which were used for the crossmatch.

\subsection{Selection filters}
We applied quality filters to discard the poorest data, adapting the criteria proposed by \cite{SOS} and following the documentation of each survey\footnote{\url{https://www.galah-survey.org/dr4/using_the_data/}}\footnote{\url{https://www.sdss4.org/dr17/irspec/use-radial-velocities/}} \citep{RAVEDR6_Steinmetz_2020,LAMOSTDR1}:

\textit{RAVE}: RV zero-point correction |\texttt{cHRV}|\;<\;$10\,\textrm{km}\,\textrm{s}^{-1}$; RV error \texttt{e\_HRV}\;<\;$8\,\textrm{km}\,\textrm{s}^{-1}$; correlation coefficient \texttt{R}\;>\;10; quality flag \texttt{Qual}\;$\neq$\;1; and a signal-to-noise ratio \texttt{S\_Nm}\;>\;0, which also flags errors when negative.

\textit{GALAH}: Signal-to-noise per pixel \texttt{snr\_px\_ccd3}\;>\;30; stellar parameters quality flag \texttt{flag\_sp}\;=\;0; and RV error \texttt{e\_rv\_comp\_1}\;<\;$100\,\textrm{km}\,\textrm{s}^{-1}$).

\textit{APOGEE}: Quality flag $\texttt{STARFLAG}\mod 2^\textrm{flag}$$\:\neq\:$0, (keeping flags 0,3,4,9,12,13,19,22); signal-to-noise ratio \texttt{SNR}\;>\;5; and RV error $0\,\textrm{km}\,\textrm{s}^{-1}$ < \texttt{VRELERR} < $2\,\textrm{km}\,\textrm{s}^{-1}$.

\textit{LAMOST-LRS}: Signal-to-noise ratio \texttt{snrg}\;>\;15; and RV error \texttt{rv\_err}\;>\;$0\,\textrm{km}\,\textrm{s}^{-1}$.         

\textit{LAMOST-MRS}: Signal-to-noise ratio \texttt{snr}\;>\;15; and RV error $0\,\textrm{km}\,\textrm{s}^{-1}$<\texttt{rv\_lasp1\_err}.\;<\;$100\,\textrm{km}\,\textrm{s}^{-1}$). 

\subsection{Source catalogues \& \textit{visit} catalogues}
The previously mentioned surveys conduct multiple observations of the same object to enhance result accuracy. Depending on the curation strategy, these are either merged into a single entry per source (hereafter referred to as \textit{source catalogue}) or stored as multiple observations (hereafter referred to as \textit{visit catalogue}). In our science case, we have to use the \textit{source catalogue}, to crossmatch between sources.

\textit{Gaia} releases \textit{source catalogues}, whereas \textit{LAMOST-LRS}, \textit{LAMOST-MRS} and \textit{RAVE} provide \textit{visit catalogues}. \textit{GALAH} and \textit{APOGEE}, offer both types of catalogues. Following each survey’s recommendations, we used the \textit{APOGEE} \textit{visit catalogue} and the \textit{GALAH} \textit{source catalogue}.

Where necessary, we transformed \textit{visit catalogues} into \textit{source catalogues}. For each star, we computed the median RV from its \textit{visits}; if the number of \textit{visits} was even, we selected the value with the smaller associated error among the two median values. When multiple RV values were available from different surveys, we retained the one with the smallest error.\\

\new{The main \textit{Gaia} DR3 table provides astrometric and RV data obtained by fitting the standard single-star model \citep{perryman1997hipparcos} to the spacecraft measurements. As a first attempt to account for binaries, \textit{Gaia} DR3 also includes dedicated tables that refit poor astrometric results as accelerated models compatible with binaries \citep{GDR3AstroBinStarProc_Halbwachs,GDR3AstrobinMC_Holl}, and variable RVs as eclipsing or spectroscopic binaries \citep{GDR3_SB1_Gosset}. When available, we replaced the original astrometric or RV entries with the corresponding binary solutions.}

\new{As with astrometry, RV data cannot be used directly from the archive tables without correction for known systematics. For cool stars (\texttt{rv\_template\_teff} < 8500 \textrm{K}), we applied the prescription of \cite{GaiaDR3_RV}, for hot stars, that of \cite{GDR3HotRV}; and for the uncertainties, the recipe of \cite{GDR3CatalogueVerif}.}

The final sample includes \new{$99\,416$} sources within 80 pc of $\beta$ Pic with RVs and associated uncertainties\textemdash representing \new{63.3\%} of the total GCNS sources in this volume. The remaining \new{57\,579 (36.7\%)} lack RV data. The distribution of RV sources across surveys is summarised in Table \ref{tab:rvdat}. Note that the \textit{Gaia} counts differ from those mentioned earlier in this Section, as we adopted RV data from the other surveys when they offered lower uncertainties.
\new{\subsection{Convective blueshift and gravitational redshift}
Spectroscopic radial velocities should not be used as kinematic radial velocities \citep{FundDefRV_Lindegreen}. Two main effects contribute to systematic shifts in the measured line positions: the gravitational redshift and the convective blueshift.}

\new{ The gravitational redshift arises because photons lose energy when escaping from the stellar gravitational potential of the star, producing a systematic redshift that depends on the stellar mass and radius (see \citealt{Kages_Couture_2023}), but in this work we adopted mass and $\log g$ due to data availability:}

\begin{equation}
    \new{\Delta \textrm{RV}_{\textrm{grav}}=c\left(\left( 1-\dfrac{2\sqrt{GM\times10^{\log g}}}{c^2}\right)^{-1/2}-1\right)}
\end{equation}

\new{The convective blueshift is related to the dynamics of convective atmospheres:  hotter, rising gas contributes more flux than cooler, descending material, leading to asymmetric spectral lines and a net blueshift. Its amplitude depends on the stellar structure\textemdash being significant in giants and cool dwarfs with outer convective layers, but negligible in hot dwarfs and white dwarfs. For the corrections, we followed the formulation of \cite{Kages_Couture_2023} for dwarfs, and the dependence described by \cite{LiebingBlueshiftGiants} for evolved stars.} \new{Typical values of the gravitational redshift are 0.4-0.6 $\textrm{km\,s}^{-1}$, whereas the convective blueshift generally has lower absolute values, in the range 0.02-0.4 $\textrm{km\,s}^{-1}$.}\\

\new{As a source of stellar parameters, we used the same spectroscopic surveys that provided RV data, complemented with the TESS Input Catalogue (TIC, \citealt{Paegert2021arXiv}) values, and additional works dedicated to white dwarfs \citep{Gianninas2015,Bedard2017,Killic2020,Bonavita2020,JimenezEsteban2023,Vincent2024}. The TIC catalogue classifies stars as dwarfs (main-sequence and white dwarfs), giants, or subgiants (the latter two being evolved stars). We also employed the luminosity classes of spectral types available in SIMBAD to complete the classification; objects without spectral type information were assumed to be dwarfs. Once classified, missing parameters were estimated by interpolation within each class: for evolved stars and white dwarfs we used available data for objects with similar known parameters, while for main-sequence dwarfs, we relied on the updated version\footnote{\url{https://www.pas.rochester.edu/~emamajek/EEM_dwarf_UBVIJHK_colors_Teff.txt}} of Table 3 of \citet{PMamajek2013}. Objects with no available data were assumed to be faint, low-mass stars and were assigned the corresponding gravitational redshift and convective blueshift.}

\new{After applying the corrections to all sources, the kinematic radial velocities can be used.} The complete catalogue is available at \href{https://zenodo.org/records/15374714?preview=1&token=eyJhbGciOiJIUzUxMiJ9.eyJpZCI6ImFkNDJiMTMzLTY3NDQtNDliMy1iYzAyLWYyZDhjZjZjZjRlOSIsImRhdGEiOnt9LCJyYW5kb20iOiI2ZGJjMWE3ZWM1YTYyYjM4Nzg2NmVlZmIwZmE0NzRkYSJ9.IRWx0bKERevZqB6a9pRSl4M_dcKTOAUC3g0gf2LsLBAxlEer7VPuFWk4LZvFRnauYwJApsXp8BouHz7XJ-yCTQ}{Zenodo}.

\begin{table}

    \caption{\new{Radial velocity data for the GCNS sources within 80 pc of $\beta$ Pic. The first column lists the origin of the RVs, and the second column lists the corresponding number of stars.}}
    \label{tab:rvdat}
    \centering
    \begin{tabular}{l r}
        \toprule
        \textbf{Catalogue} & \textbf{Number of stars} \\
        \midrule
        
        GDR3                  & 87\,066 \\
        \textit{APOGEE}       & 6\,570  \\
        GCNS                  & 2\,921  \\
        GDR3-NSS              & 1\,866  \\
        \textit{RAVE}         & 486     \\
        \textit{GALAH}        & 404     \\
        \textit{LAMOST-MRS}   & 66      \\
        \textit{LAMOST-LRS}   & 37      \\
        \midrule
        \textbf{TOTAL} & \textbf{99\,416} \\ 
        \bottomrule
    \end{tabular}

\end{table}

\begin{figure*}[ht]

\centering
\includegraphics[width=0.95\textwidth]{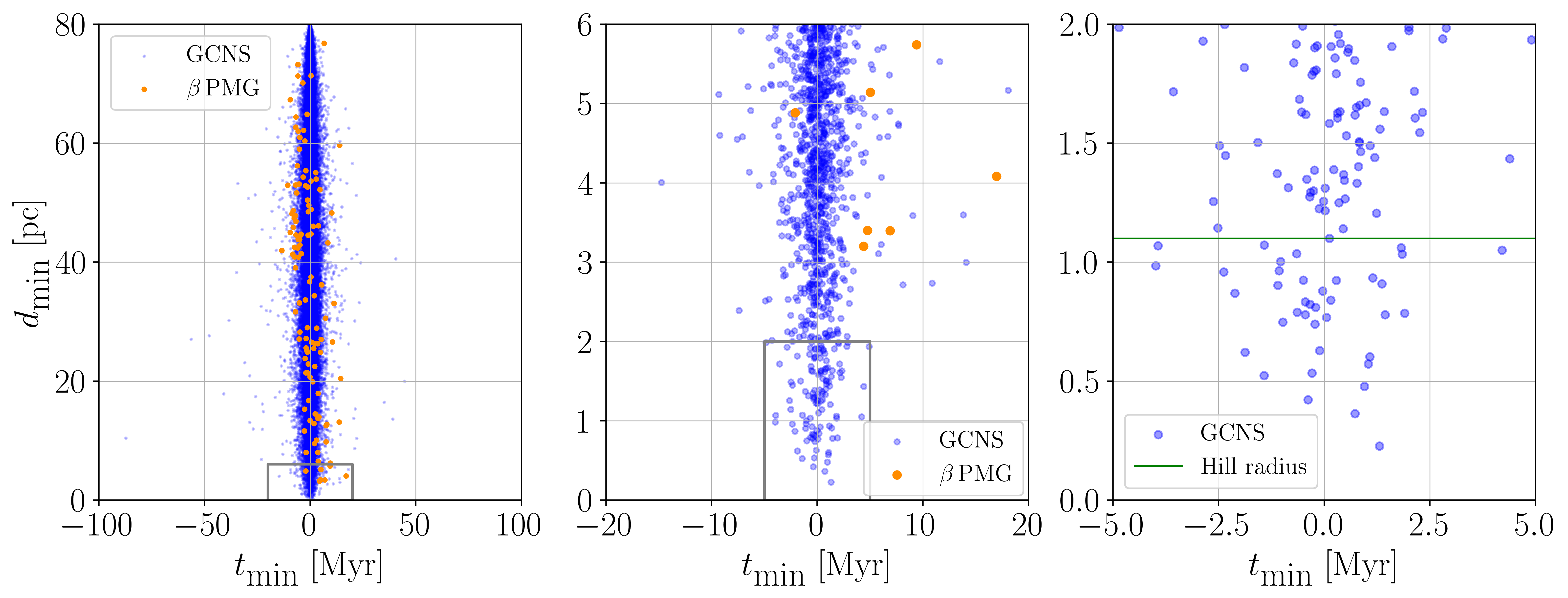}
\caption{Distance versus time distribution of each close encounter of our sample of stars with $\beta$ Pictoris (see Table \ref{tab:rvdat}). The first panel shows the full sample of \renew{$99\,416$} from the GCNS catalogue. The second panel displays the subset of \renew{1\,005} within 6 pc. The third panel highlights the \renew{116} stars with a closest approach distance $d_{\textrm{min}}<2\;\textrm{pc}$. The members of the $\beta$ Pictoris Moving Group are shown in orange.}
\label{fig:tdsingle}
\end{figure*}

\section{Close encounters with \texorpdfstring{$\beta$}{beta} Pictoris}
\label{sec:CatalogueEncounters}

In order to calculate the stellar encounters with $\beta$ Pic and its nearby stars, we trace their orbits backwards and forwards in time and calculate the minimum distance between their positions over time.


In these integrations we neglect star-star interactions \new{(see discussion in Section \ref{sub:errors})}, so each orbit is determined solely by its initial conditions and the Galactic potential.
We used the Galactic potential \texttt{MilkyWayPotential2022} presented in the Python package \texttt{Gala} \citep{galapaper,galacode} to perform the integration of stellar motions within a Galaxy model. This potential comprises four components: a Hernquist bulge and nucleus, a Miyamoto-Nagai disc, and a Navarro-Frenk-White dark matter halo. The developers fitted the disk model to the \cite{MWRotCurve} rotation curve for radial dependence, and its vertical structure was fit to the phase-space spiral in the solar neighbourhood as described in \cite{MW2022}\footnote{\url{https://gala-astro.readthedocs.io/en/latest/api/gala.potential.potential.MilkyWayPotential2022.html}}. \new{The parameters defining the potential are listed in Table \ref{tab:galpo}.}

\begin{table}[h!]

\centering
\caption{\new{Galactic potential parameters of \texttt{MilkyWayPotential2022}.}}
\label{tab:galpo}
\begin{tabular}{llr}
\hline\hline
Component & Parameters & Values \\
\hline
\noalign{\smallskip}  
Hernquist bulge & Mass & $5 \times 10^{9}\,M_\odot$ \\
                & Core radius & $1.0$ kpc \\
Hernquist nucleus & Mass & $1.8142 \times 10^{9}\,M_\odot$ \\
                  & Core radius & $0.068887$ kpc \\
Disc (3 $\times$  & Total mass & $4.7717 \times 10^{10}\,M_\odot$ \\
Miyamoto--Nagai)                & Radial scale length & $2.6$ kpc \\
                & Vertical scale height & $0.3$ kpc \\
NFW halo & Scale mass & $5.5427 \times 10^{11}\,M_\odot$ \\
         & Scale radius & $15.626$ kpc \\
\hline
\end{tabular}
\end{table}

Prior to integration, we transformed the ICRS \textit{Gaia} coordinates into a Galactocentric reference frame using \texttt{Astropy}, \new{adopting the following parameters: the ICRS position of the Galactic Centre at $[\textrm{RA, DEC}]_{\rm GC}=[17^{\rm h}45^{\rm m}37.224^{\rm s},-28^\circ56'10.23''$ \citep{Reid_2004}, the distance from the Sun to the Galactic Centre of 8.275\,kpc \citep{GalactocentricdistNueva}, the Sun's height above the Galactic plane $z_{\odot}=20.8\ \mathrm{pc}$  \citep{SunVelocityGalactNueva}, and its Galactocentric velocity $[U_\odot, V_\odot, W_\odot]=[8.4, 251.8, 8.4]\:\textrm{km\,s}^{-1}$ \citep{SunVelocityGalactNueva}.}

We integrated the orbits of $\beta$ Pic and the remaining \new{$99\,416$} stars both forwards and backwards in time using a timestep of $|\Delta t|=50\,\textrm{yr}$ until a maximum integration time $t_\textrm{max}$. 
\new{Because \texttt{Gala} requires the integration interval to be defined in advance, halting and restarting the simulation at each timestep to check if the minimum was reached may introduce cumulative numerical errors. For both numerical safety and efficiency, we therefore adopted a sequence of integration windows from the present to} $|t_\textrm{max}|=[5,10,15,20,25,30,35,40,60,100]~\textrm{Myr}$, checking at each whether the minimum Euclidean distance between $\beta$ Pic and a given star was achieved. If so, the integration terminated; otherwise, it continued to the next $t_\textrm{max}$. Though integrations beyond $\beta$ Pic's 20 Myr lifespan are not physically meaningful, they are useful for statistical analysis. Specific Energy conservation during integration was verified to machine precision, $\textrm{max}(|\Delta E/E_0|)<10^{-14}$.

\begin{figure}[ht]

\centering
\includegraphics[width=0.4\textwidth]{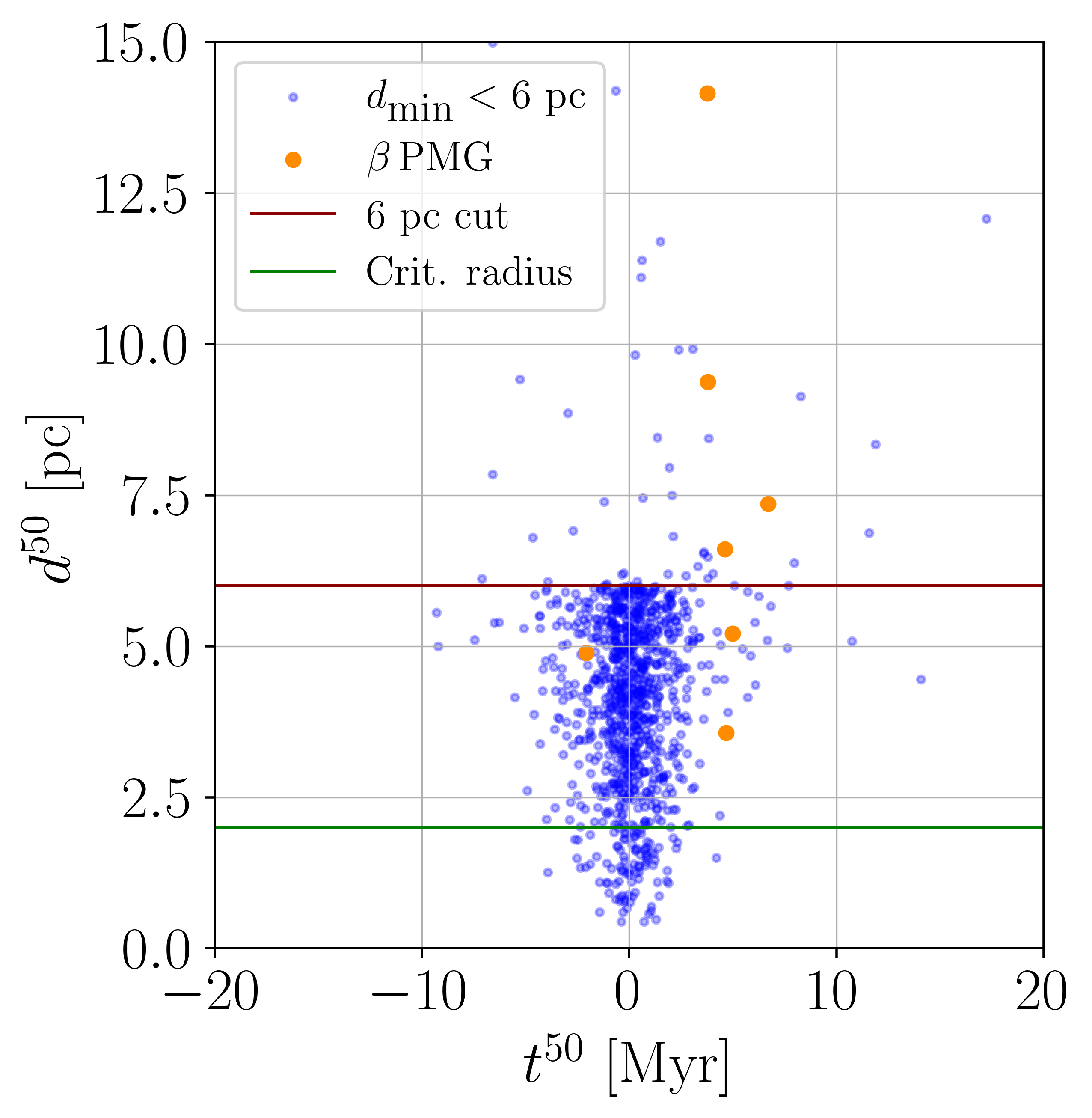}
\includegraphics[width=0.4\textwidth]{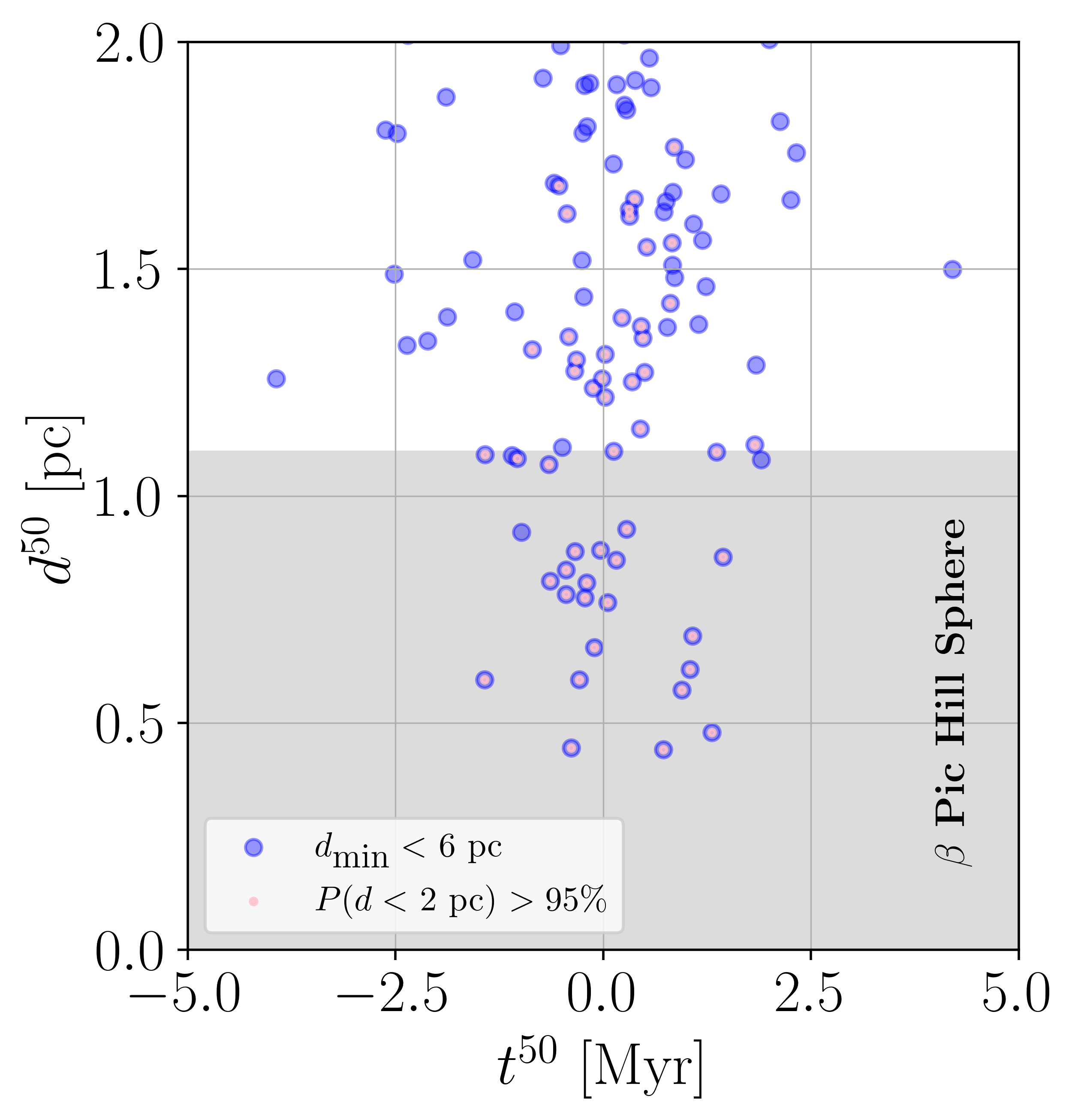} %
\caption{Upper: Medians of the distance versus time distributions for the \renew{1\,005} candidate encounters with $\beta$ Pic that had $d_\textrm{min}<6$ pc, calculated using the method described in Section \ref{sub:errors}. The horizontal red line marks the 6 pc limit, used to select these sources based on their $d_\textrm{min}$ values (second panel of figure \ref{fig:tdsingle}). Distributions with $d^{50}$ values above this line correspond to the \renew{47} more distant, higher-dispersion encounters. The horizontal green line denotes the 2 pc \textit{critical radius}. The orange points indicate the $\beta\,$PMG members. Lower: same as above, but for the \renew{96} sources with $d^\textrm{50}<2\,\textrm{pc}$. The Hill sphere of $\beta$ Pic is highlighted in grey. Only \renew{49} stars have at least a 95\% empirical probability of passing within 2 pc of $\beta$ Pic, and are considered actual encounters (highlighted in pink).}
\label{fig:tcdc993}
\end{figure}

We used the fourth-order symplectic integrator of \cite{Ruth4th}, as implemented in \texttt{Gala}. Each integrated orbit is represented by galactocentric coordinates ${X_i,\,Y_i,\,Z_i}$ and velocities ${U_i,\,V_i,\,W_i}$ along a time vector ${t_i}$. After integration, we computed the Euclidean distance $d_i$ between $\beta$ Pic and each star at each $t_i$, identifying the time $t_\textrm{min}$ and distance $d_\textrm{min}=\min d_i$ of closest approach. Relative encounter velocities were also stored.

We are only interested in stars that approach closely enough to affect the surroundings of $\beta$ Pic significantly. The particles gravitationally bound to $\beta$ Pic must lie within its Hill radius in the gravitational potential of the Galaxy, which is about 1.1 pc \citep{kalas2001stellar}. To estimate the distance at which a stellar flyby significantly influences the system's exocomets, we employ the concept of the \textit{Critical radius} \citep{torres2019}. Using a particle from the cloud of comets of $\beta$ Pic with a semi-major axis of $a=10^5$ au and an eccentricity $e=0.9$, we estimated this radius to be $d_\textrm{crit}=2\;\textrm{pc}$, corresponding to an impulse equal to one-thousandth the orbital velocity of the particle at apocentre.

In Figure \ref{fig:tdsingle}, we show the distribution of closest distances $d_{\textrm{min}}$ as function of encounter times $t_{\textrm{min}}$. As integration time increases, the number of encounters \new{decreases. The stars of the present-day sample typically have encounters in recent times and then move further away from $\beta$ Pic. To identify additional encounters, we would need to increase the present-day sample; however, beyond $100\,\textrm{pc}$ from the Sun, the sample completeness across spectral types drops rapidly \citep{GCNS}.}
One might expect a time asymmetry due to $\beta$ Pic's youth ($\sim$20 Myr), yet no such effect was observed in the full sample. However, when examining $\beta\,$PMG members using the \cite{BPMG_Luhman_2024} census, more past than future encounters were found, although none were among the closest. Indeed, none of the closest encounters studied later on are with $\beta\,$PMG members. 



\begin{figure}[ht]
\centering
\includegraphics[width=0.45\textwidth]{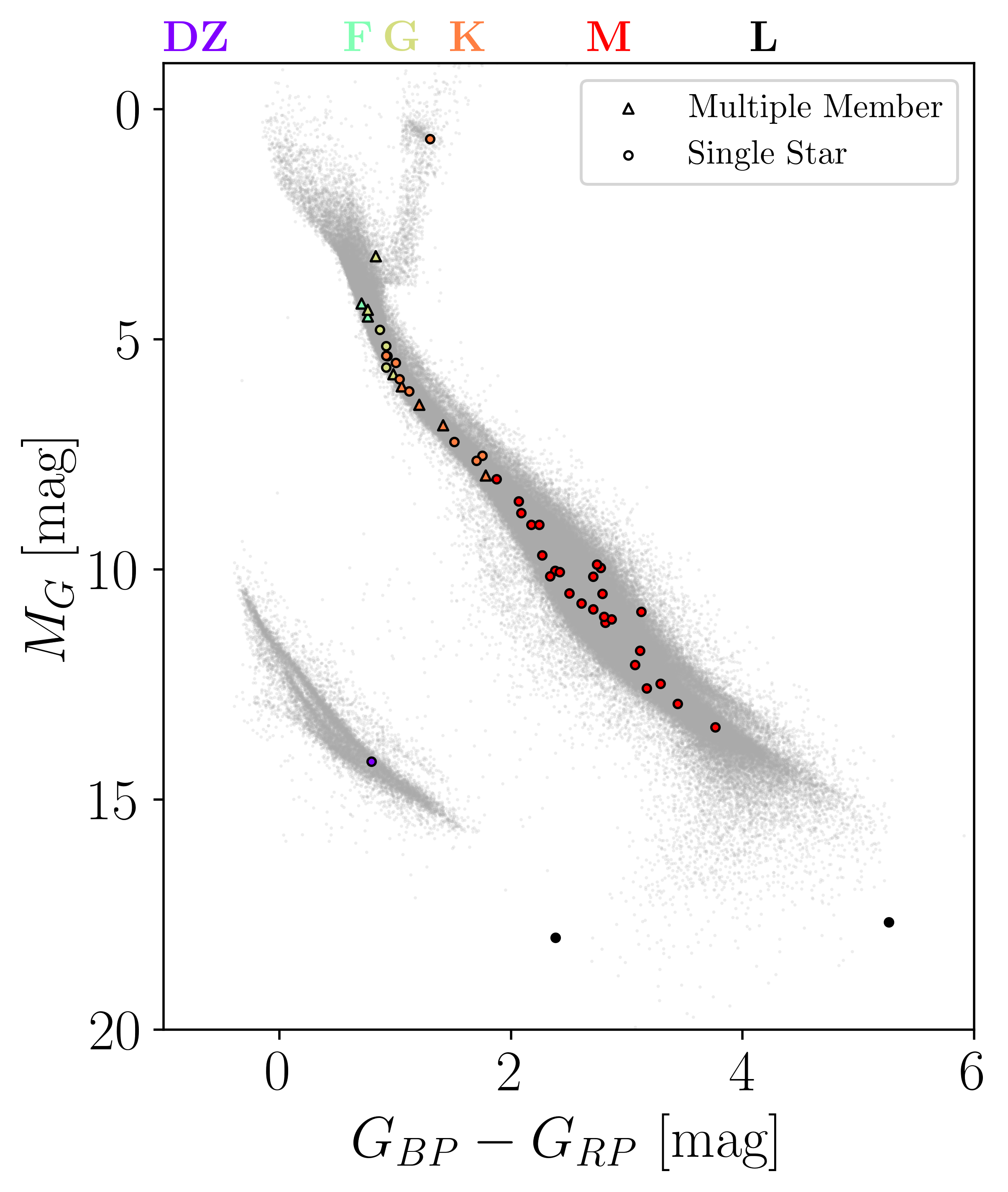}
\caption{\textit{Gaia} colour-magnitude Hertzsprung-Russell diagram for the 52 stars that present a close encounter with $\beta$ Pic (see Table \ref{tab:encounters}). The grey field sources were obtained from a \textit{Gaia} query using magnitude-over-error thresholds matching the lowest-quality values among the encounter sources: \texttt{parallax}>10 mas (maximum distance of 100 pc), \texttt{parallax\_over\_error}>58, \texttt{phot\_g\_mean\_flux\_over\_error}>220, \texttt{phot\_bp\_mean\_flux\_over\_error}>2.3, and \texttt{phot\_rp\_mean\_flux\_over\_error}>23.5. Dot colours represent the spectral types.}

\label{fig:HR}
\end{figure}

\subsection{Uncertainty estimation}
\label{sub:errors}
We used a Monte Carlo method to assess uncertainties for the closest star encounters (<6 pc) with $10\,000$ orbit integrations per star. To manage computational cost, we applied a 6 pc preselection cut, reducing the sample to \new{1\,005} stars. This threshold, discussed later, balances feasibility with the risk of missing significant close approaches.




For each of the selected stars (including $\beta$ Pic), we generated $10\,000$ samples \new{(hereafter referred to as \textit{clones})} from a 6D Gaussian distribution defined by GDR3 astrometric and RV data, incorporating full covariance matrices. For each \new{clone} $j$ of a given star, we computed the encounter parameters \new{with the corresponding $j$ clone of $\beta$ Pic}, using the same orbital integration method described previously. \new{We stored the three components of the relative Galactocentric position ($X_j, Y_j, Z_j$) and velocity ($U_j, V_j, W_j$) with respect to $\beta$ Pic, and their modules, as well as the encounter time $t_i$.} 

Figure \ref{fig:tcdc993} displays the medians ($50^\textrm{th}$ percentiles) of the marginal distributions of encounter times ($t^{50}$) and distances ($d^{50}$) for these stars. As in Figure \ref{fig:tdsingle}, there is a noticeable clustering of encounters near the present epoch. We identify \new{47} stars with $d^{50}> 6\;\textrm{pc}$, indicating large uncertainties and highly scattered distance distributions. 

To define a high-confidence subset of encounters, we retained only stars with $>95\%$ empirical probability (i.e., $>9\,500$ out of $10\,000$ \new{clones}) of passing within 2 pc. This criterion yielded \new{49} stars that will encounter $\beta$ Pic (see Table \ref{tab:encounters}). A more detailed assessment of the 95\% threshold is presented in Section \ref{subsub:pt95}. Notably, all of these have closest-approach distances in the single-orbit integration below \renew{1.8 pc}, validating our initial 6 pc preselection cut. In Section \ref{sub:mul}, we assess the multiplicity of the encounters. As can be seen in the \textit{Gaia} colour-magnitude Hertzsprung-Russell diagram of the sources shown in Figure \ref{fig:HR}, the majority of the stars are main-sequence FGKM dwarf stars, except the red giant GDR3 2887731882922767744, the subgiant GDR3 6353376831270492800, the L-type brown dwarfs GDR3 5185493447310441728 and GDR3 1311454726097258368, \new{and the white dwarf GDR3 1193520666521113344}.

\renew{To visualise the clone distribution in the $(X,Y,Z,U,V,W,t)$ encounter parameter space, we computed the Mahalanobis distance of each clone relative to the overall distribution and retained the 50\% with the lowest values as the most clustered subset. An example corner plot is shown in Figure \ref{fig:corner}. This subset contains the strongest perturbers. The expected hyperbolic deviation from a straight-line path was also estimated and found to be $<0.01\,\textrm{pc}$ for most of the clones, supporting the neglect of star–star interactions in the integrations. Corner plots for all other encounters are provided at \href{https://zenodo.org/records/17177886?preview=1&token=eyJhbGciOiJIUzUxMiJ9.eyJpZCI6IjQ0N2Y0MjdhLTVhNWQtNGIwZC04YWE5LWU3OWU3NGJkNmMzZSIsImRhdGEiOnt9LCJyYW5kb20iOiJlYTA5ZGE4YWRiNjE1NGVkYjZjMjg4YzU0MDg5ZTcwYiJ9.MdZxvxm5K1lp-G_qTQE-1uinkYpMAUVDumVeYZbVPfm_2L4pMuSj2fyG0MZEM5Z96DQ1CRm21-5FALU7jRBf9g}{Zenodo}.}

\renew{Table \ref{tab:encounters} reports the marginal 95\% confidence intervals for the scalar magnitudes of the high-confidence encounters and their companions. For extremely close encounters, the ones with $d^{2.5}<0.1\,\textrm{pc}$ one-sided intervals were adopted; otherwise, two-sided intervals were used.}

\subsection{Binary Systems}
\label{sub:mul}
\new{In the previous section, we have treated all the stars as single when propagating their orbits. To account for multiplicity, we searched for companions in} the Washington Double Star Catalogue (WDS, \citealt{WDS}), the Multiple Star Catalogue (MSC, \citealt{MSC}), and the wide binary candidate list from \cite{ElBadry} based on GEDR3 astrometry. \new{We identified 8 stars among our 49 close encounters samples as members of 5 binary systems.} These cases require further scrutiny to accurately assess their encounter parameters and the possible influence on the $\beta$ Pic system. Multiplicity can significantly enhance the perturbative effect\textemdash sometimes by more than an order of magnitude\textemdash potentially leading to outcomes such as pair disruptions, stellar collisions, or member captures \citep{binaryencounters,Torres2023}. 

\new{For 4 of the binary systems, full astrometric and RV data are available for both components. One additional companion, GDR3 6377398274119547392 to GDR3 6353376831270492800, lacks RV data. We also added to Table \ref{tab:encounters} the companion GDR3 4078432297860547072 of GDR3 4078432504018987904, which as a single star has an empirical encounter probability with $\beta$ Pic of less than 87\%.} 

\new{For binaries with complete data for both components, we drew stellar masses for each clone from a normal distribution defined by the reported value and its uncertainty, computed the center-of-mass positions and velocities, and integrated their orbits accordingly. The resulting encounter parameters are listed in Table \ref{tab:binary}. As with the single-star calculations, the center-of-mass trajectories confirm the possibility of a close encounter of these systems with $\beta$ Pic.}\\

\begin{figure*}[ht]

\centering
\includegraphics[width=0.95\textwidth]{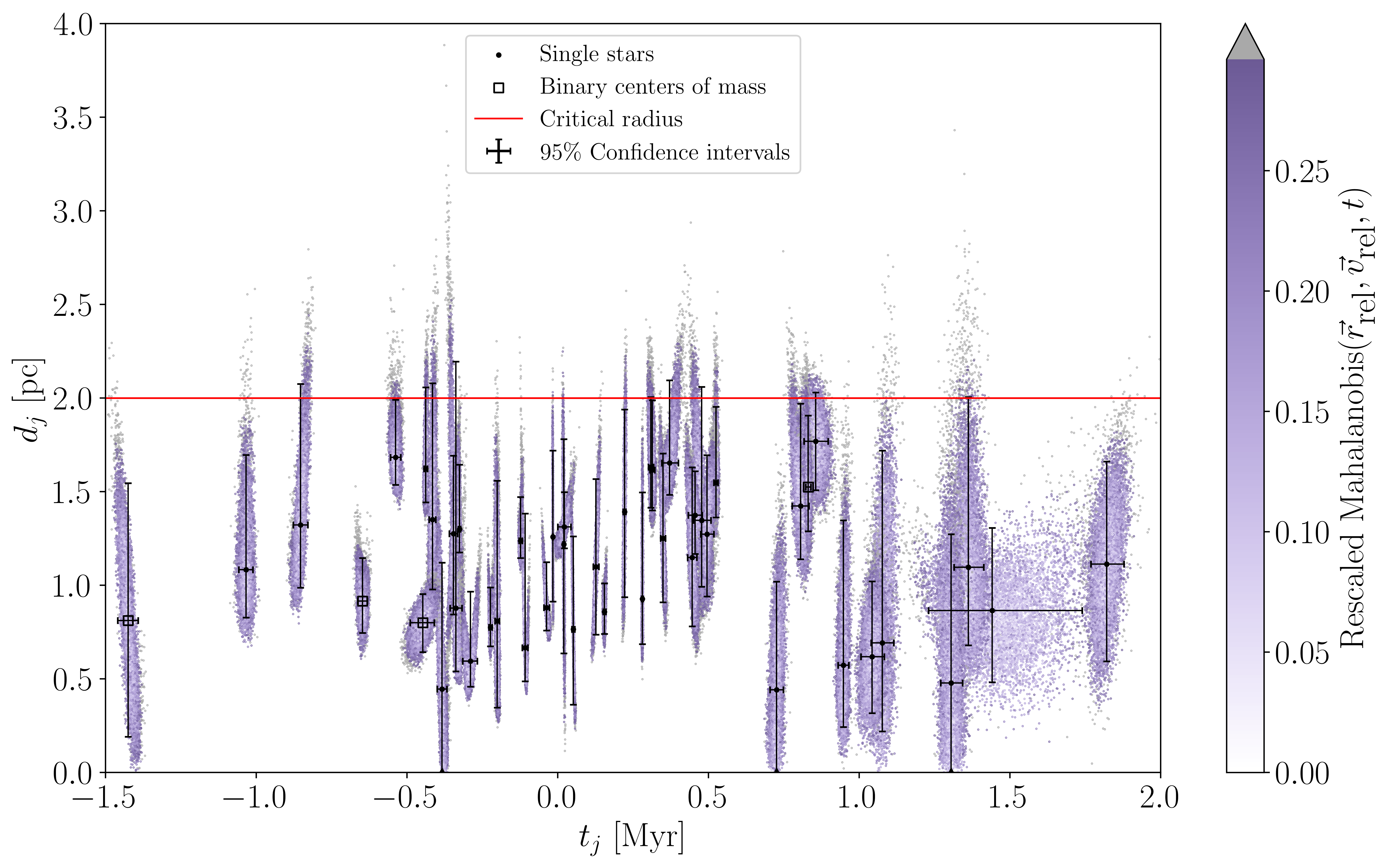}
\caption{Distance versus time distributions for the \renew{49} encounters with $\beta$ Pic that have an empirical probability of at least 95\% of occurring within 2 pc.
 The coloured points indicate the rescaled Mahalanobis distances (ensuring a common colour scale for all stars) derived from the encounter parameters. Points above the $95^\textrm{th}$ percentile of this metric are shown in grey. The black bars represent the $95\%$ confidence intervals in time and distance, obtained from the $2.5^\textrm{th}$ and $97.5^\textrm{th}$ percentiles of the corresponding marginal distributions for the further encounters and using the $95^\textrm{th}$ percentile to define the one-tailed distance interval for the closest ones. The central points are the  $50^\textrm{th}$ percentiles. The \textit{critical radius} of 2 pc is indicated by the red horizontal line. The centres of mass of binary systems, listed in table \ref{tab:binary}, are marked at the pair [$t^{50},d^{50}$] corresponding to that star.}
\label{fig:encounters51}
\end{figure*}

Figure \ref{fig:encounters51} shows the distributions of closest-approach distances and times for the final subset, \new{using the centres of mass for the binary systems}. As previously noted and visible in Figure \ref{fig:encounters51}, the encounter rate is highest near the present epoch and declines steeply with time (see Section \ref{sub:compl}).

\subsection{Encounter completeness}
\label{sub:compl}
We acknowledge that our encounter catalogue (Table \ref{tab:encounters}) constitutes a lower limit on the true number of close encounters experienced by $\beta$ Pic. As shown in Table \ref{tab:rvdat}, approximately one-third of the stars currently known in the vicinity of $\beta$ Pic lack RV measurements, making orbital reconstruction for these stars impossible.

Our method identifies encounters by integrating stellar orbits and detecting passages within a critical distance of $\beta$ Pic. Given the typical relative velocities of the encounters listed in Table \ref{tab:encounters}, a star can traverse the initial 80 pc sphere around $\beta$ Pic in just a few Myr or less. Consequently, the number of detected encounters drops sharply when looking further backward or forward in time. Beyond $\sim$2 Myr, the stellar encounters drop significantly\textemdash a trend clearly illustrated in Figures \ref{fig:tdsingle}, \ref{fig:tcdc993}, and \ref{fig:encounters51}. 

Expanding the search volume by increasing the initial radius would significantly improve this limitation. For example, of the \renew{49} high-confidence encounters listed in Table \ref{tab:encounters}, only \renew{7} are presently located at distances greater than 50 pc from $\beta$ Pic, and all but one of these occurred outside the $\pm$0.5 Myr interval. However, the stars currently located further away would yield less accurate encounters because of the limitations of \textit{Gaia} data.


\begin{table*}
\centering

\caption{Encounter parameters of the 49 stars with a 95\% empirical probability of having a close encounter ($d<2\;\textrm{pc}$) with $\beta$ Pictoris, or their companions, treating all sources as single. Columns list the \textit{Gaia} DR3 ID, spectral type, stellar mass, and the encounter parameters (time, distance, and velocity at pericentre). Values are given as confidence intervals: $x^{50}$ and $[x^{2.5}, x^{97.5}]$ for $x\in\{t,d,v\}$, and $[0, d^{95}]$ for extremely close encounters. This table is available in \href{https://zenodo.org/records/15442557?preview=1&token=eyJhbGciOiJIUzUxMiJ9.eyJpZCI6IjYxM2Y2ZTYzLTQxOWQtNGIyOC1hYjIzLWZlYmMzMDMwYzZjZiIsImRhdGEiOnt9LCJyYW5kb20iOiI4YjJmMTdkOWMyMGRiYWVlY2QxM2Q4MDhhNDUzOWVkNCJ9.aV3EwuKSwFFaedlbsFVv-BHk9fP0h_cP6PEM2_KikvTV8XgOmwCm2vsYKQQxgIM7bnnR7SqZoiC81KSAfDwuxA}{Zenodo}.}


\begin{tabular}{clccccccc}

\hline\hline
\noalign{\smallskip}

        \multirow{2}{*}{\textit{Gaia} DR3 ID} & \multirow{2}{*}{Spectral} & \multirow{2}{*}{Mass} & 
        \multicolumn{2}{c}{Enc. Time [Myr]} & 
        \multicolumn{2}{c}{Enc. Distance [pc]} & 
        \multicolumn{2}{c}{Enc. Velocity [km s$^{-1}$]} \\ 
        
        \cmidrule(lr){4-5} \cmidrule(lr){6-7} \cmidrule(lr){8-9}
        
       & Type & [$M_\odot$] & $t^{50}$ 
 & [$t^{2.5}$, $t^{97.5}$] & $d^\textrm{50}$ 
 & [$d^{2.5},d^{97.5}]|[0,d^{95}]$ &  $v^\textrm{50}$ 
 & [$v^\textrm{2.5}$, $v^\textrm{97.5}$] \\

\hline
\noalign{\smallskip}

4841460210350512768\tablefootmark{1}\;\; &  K     & 0.75  & -1.428 & [-1.469, -1.390] & 0.596 & [0.194, 1.196] & 42.52 & [41.39, 43.63] \\
4841460279069516672\tablefootmark{1}\;\; & F8V    & 1.05  & -1.423 & [-1.458, -1.389] & 1.091 & [0.358, 1.840] & 42.59 & [41.64, 43.54] \\
6034774508013700992\;\;\;\;              & K      & 0.67  & -1.034 & [-1.058, -1.011] & 1.083 & [0.827, 1.696] & 48.78 & [47.69, 49.87] \\
6353376831270492800\tablefootmark{1,2}   & G2IV-V & 1.12  & -0.853 & [-0.877, -0.829] & 1.323 & [0.988, 2.075] & 23.92 & [23.35, 24.49] \\
3712538811193759744\tablefootmark{1}\;\; & K2V    & 0.83  & -0.655 & [-0.672, -0.638] & 1.070 & [0.897, 1.283] & 43.54 & [42.65, 44.44] \\
3712538708114516736\tablefootmark{1}\;\; & G9V    & 0.87  & -0.641 & [-0.658, -0.625] & 0.813 & [0.611, 1.178] & 44.49 & [43.55, 45.43] \\
3173680600645518848\;\;\;\;              &  M4V   & 0.42  & -0.538 & [-0.555, -0.521] & 1.684 & [1.537, 1.991] & 28.86 & [28.02, 29.78] \\
4763897739549071744\tablefootmark{1}\;\; &  K7V   & 0.58  & -0.450 & [-0.497, -0.410] & 0.838 & [0.686, 0.989] & 17.87 & [16.66, 19.09] \\
4763906879239461632\tablefootmark{1,2}   &  F9V   & 1.13  & -0.447 & [-0.488, -0.408] & 0.783 & [0.621, 0.941] & 18.02 & [17.00, 19.09] \\
5850123968410499200\;\;\;\;              &  K     & 0.63  & -0.438 & [-0.444, -0.433] & 1.622 & [1.443, 2.058] & 64.60 & [63.81, 65.39] \\
6453242173886773376\;\;\;\;              &  M     & 0.26  & -0.415 & [-0.427, -0.405] & 1.351 & [0.979, 2.079] & 53.69 & [52.35, 55.05] \\
2898608526823037184\;\;\;\;              &  M3.5  & 0.38  & -0.384 & [-0.400, -0.368] & 0.446 & [\;\;\;\;\;\;0, 1.121] & 30.83 & [29.57, 32.17] \\
5117974602912370432\tablefootmark{2}\;\; &  G8V   & 0.93  & -0.346 & [-0.360, -0.333] & 1.275 & [0.845, 1.693] & 40.15 & [39.39, 40.92] \\
5796958595407301632\;\;\;\;              &  M     & 0.16  & -0.338 & [-0.358, -0.318] & 0.878 & [0.541, 2.196] & 72.41 & [68.17, 76.94] \\
6370513647703084544\;\;\;\;              &   G9V  & 0.90  & -0.326 & [-0.330, -0.323] & 1.300 & [1.175, 1.645] & 52.28 & [51.78, 52.79] \\
5296211588171838720\;\;\;\;              &   M    & 0.21  & -0.289 & [-0.315, -0.267] & 0.595 & [0.459, 0.967] & 59.31 & [54.65, 63.89] \\
5540934254857896192\;\;\;\;              &  M     & 0.34  & -0.223 & [-0.229, -0.218] & 0.776 & [0.674, 0.989] & 76.04 & [74.83, 77.25] \\
2954555801611979648\;\;\;\;              &  M5.5V & 0.13  & -0.201 & [-0.208, -0.194] & 0.809 & [0.347, 1.559] & 37.12 & [36.65, 37.61] \\
4759543295545939840\tablefootmark{2}\;\; & G5V    & 0.91  & -0.123 & [-0.128, -0.118] & 1.238 & [1.146, 1.473] & 24.51 & [24.21, 24.82] \\
2887731882922767744\tablefootmark{2}\;\; & K1III  & 1.70  & -0.108 & [-0.118, -0.099] & 0.666 & [0.487, 1.384] & 87.00 & [84.46, 89.47] \\
4768702571002262912\;\;\;\;              & M2     & 0.48  & -0.037 & [-0.047, -0.028] & 0.881 & [0.759, 1.124] & 44.60 & [43.77, 45.44] \\
4767716893186840320\;\;\;\;              &  M     & 0.32  & -0.016 & [-0.017, -0.015] & 1.258 & [0.914, 1.720] & 74.20 & [73.95, 74.47] \\
4794632903476180096\;\;\;\;              &  K7V   & 0.64  &  0.020 & [\,\,0.017,\;\;0.023] & 1.218 & [0.637, 1.781] & 38.99 & [38.71, 39.30] \\
5553110654636730496\;\;\;\;              & M1     & 0.51  &  0.022 & [\,\,0.001,\;\;0.043] & 1.313 & [1.198, 1.498] & 20.77 & [19.86, 21.70] \\
4803556711646531840\;\;\;\;              &  M     & 0.29  &  0.052 & [\,\,0.047,\;\;0.057] & 0.766 & [0.363, 1.261] & 58.46 & [57.70, 59.25] \\
4757687388639045504\;\;\;\;              &  M4.5  & 0.17  &  0.127 & [\,\,0.118,\;\;0.136] & 1.099 & [0.737, 1.568] & 32.12 & [31.49, 32.81] \\
5185493447310441728\;\;\;\;              &  L3    & 0.15  &  0.155 & [\,\,0.150,\;\;0.159] & 0.859 & [0.740, 1.011] & 107.5 & [106.5, 108.5] \\
5493588665684618752\;\;\;\;              & G6.5V  & 0.94  &  0.221 & [\,\,0.217,\;\;0.226] & 1.393 & [0.936, 1.939] & 27.43 & [27.07, 27.82] \\
3441134536361404928\;\;\;\;              & M3     & 0.43  &  0.280 & [\,\,0.278,\;\;0.283] & 0.927 & [0.686, 1.496] & 146.8 & [146.1, 147.4] \\
3007559370624135424\;\;\;\;              & M3V    & 0.27  &  0.309 & [\,\,0.301,\;\;0.317] & 1.632 & [1.414, 2.006] & 44.49 & [43.87, 45.11] \\
2313022171603701888\;\;\;\;              & M3V    & 0.32  &  0.313 & [\,\,0.305,\;\;0.322] & 1.616 & [1.401, 1.988] & 57.73 & [56.89, 58.59] \\
2460983348274381696\;\;\;\;              & M2V    & 0.37  &  0.349 & [\,\,0.341,\;\;0.357] & 1.252 & [0.910, 1.705] & 53.50 & [52.83, 54.17] \\
4670295730560582784\;\;\;\;              &  M     & 0.31  &  0.370 & [\,\,0.345,\;\;0.400] & 1.654 & [1.484, 2.095] & 200.3 & [186.1, 214.4] \\
5856411869205581568\;\;\;\;              &  K1V   & 0.82  &  0.446 & [\,\,0.431,\;\;0.461] & 1.148 & [0.781, 1.631] & 32.10 & [31.48, 32.72] \\
4817064138977294592\;\;\;\;              &  M     & 0.49  &  0.455 & [\,\,0.433,\;\;0.479] & 1.374 & [1.168, 1.610] & 34.06 & [32.82, 35.34] \\
2946531325238075776\;\;\;\;              &  M     & 0.24  &  0.477 & [\,\,0.450,\;\;0.508] & 1.348 & [0.993, 2.061] & 49.52 & [46.60, 52.42] \\
4525711600783788160\;\;\;\;              &  M5.5  & 0.12  &  0.495 & [\,\,0.474,\;\;0.517] & 1.273 & [0.941, 1.696] & 70.17 & [67.28, 73.01] \\
6413811006857073536\;\;\;\;              &  M1    & 0.59  &  0.524 & [\,\,0.517,\;\;0.532] & 1.549 & [1.363, 1.953] & 45.77 & [45.18, 46.35] \\
1311454726097258368\;\;\;\;              &  L3    & 0.07  &  0.726 & [\,\,0.704,\;\;0.748] & 0.442 & [\;\;\;\;\;\;0, 1.021] & 62.16 & [61.14, 63.18] \\
2940796611884222208\;\;\;\;              &  K0    & 0.81  &  0.805 & [\,\,0.777,\;\;0.833] & 1.424 & [1.139, 1.970] & 27.64 & [26.84, 28.47] \\
4078432504018987904\tablefootmark{1}\;\; & G3/5V  & 1.05  &  0.827 & [\,\,0.817,\;\;0.836] & 1.558 & [1.344, 1.852] & 83.09 & [82.34, 83.84] \\
4078432297860547072\tablefootmark{1,3}   &  K     & 0.70  &  0.837 & [\,\,0.829,\;\;0.845] & 1.669 & [1.285, 2.261] & 82.17 & [81.54, 82.81] \\
3864615459886222464\;\;\;\;              &  M4V   & 0.27  &  0.855 & [\,\,0.817,\;\;0.897] & 1.769 & [1.508, 2.030] & 23.29 & [22.40, 24.17] \\
6758141249403594112\;\;\;\;              & G8/K0V & 0.86  &  0.947 & [\,\,0.930,\;\;0.965] & 0.573 & [0.244, 1.347] & 48.76 & [47.96, 49.56] \\
1193520666521113344\;\;\;\;              & DZ     & 0.56  &  1.043 & [\,\,1.005,\;\;1.083] & 0.618 & [0.317, 1.021] & 33.48 & [32.36, 34.58] \\
5038817840251308288\tablefootmark{1}\;\; & K0V    & 0.92  &  1.076 & [\,\,1.039,\;\;1.115] & 0.692 & [0.220, 1.720] & 16.29 & [15.76, 16.82] \\
1028306773725676672\;\;\;\;              &  M     & 0.14  &  1.304 & [\,\,1.270,\;\;1.342] & 0.479 & [\;\;\;\;\;\;0, 1.274] & 35.92 & [34.94, 36.88] \\
6394330650108004992\;\;\;\;              &  M3V   & 0.54  &  1.362 & [\,\,1.314,\;\;1.413] & 1.097 & [0.679, 2.005] & 14.86 & [14.35, 15.37] \\
4776148635544170752\;\;\;\;              &  M     & 0.25  &  1.441 & [\,\,1.229,\;\;1.740] & 0.866 & [0.482, 1.306] & 39.01 & [32.36, 45.61] \\
1592423313280131200\;\;\;\;              &  M     & 0.34  &  1.821 & [\,\,1.768,\;\;1.878] & 1.114 & [0.594, 1.661] & 31.42 & [30.50, 32.30] \\

\hline

\end{tabular}
\label{tab:encounters}
\tablefoot{\\
\tablefoottext{1}{Multiple systems (see table \ref{tab:binary}).}\\
\tablefoottext{2}{Reported by \cite{kalas2001stellar} (see table \ref{tab:kalas}).}\\
\tablefoottext{3}{This star forms a binary system with GDR3 4078432504018987904.}
}
\end{table*}

\begin{table*}
\centering

\caption{Encounter parameters of the centres of mass of the binary systems in which at least one member is among the 49 stars with a 95\% empirical probability of having a close encounter with $\beta$ Pictoris. The system names were taken from the Washington Double Star catalogue (WDS, \citep{WDS}). For the group with no designated name, the association was identified in \cite{ElBadry}. 
The fourth and fifth groups of columns represent the encounter times and encounter distances, and the last column provides the projected separation of the two components of the system at the present time.}

\begin{tabular}{c cc  cc cc c}

\hline\hline
\noalign{\smallskip}
 \multirow{2}{*}{WDS Name} &  \multirow{2}{*}{\textit{Gaia} DR3 ID$_1$} &  \multirow{2}{*}{\textit{Gaia} DR3 ID$_2$} &  \multicolumn{2}{c}{Enc. Time [Myr]} &  \multicolumn{2}{c}{Enc. Distance [pc]} &   Projected  \\
 \cmidrule(lr){4-5} \cmidrule(lr){6-7}
                            &                                           &                                            &  $t^{50}$ & [$t^{2.5}$, $t^{97.5}$] & $d^{50}$  & [$d^{2.5}$, $d^{97.5}$]            &   Separation [au]\\

\hline
\noalign{\smallskip}   
J03572\,--4413 & 4841460279069516672 & 4841460210350512768 & -1.43  & [-1.46, -1.39]   & 0.81 & [0.19, 1.55] & 934  \\
J13237+0243    & 3712538811193759744 & 3712538708114516736 & -0.65  & [-0.66, -0.63]   & 0.92 & [0.75, 1.15] & 434  \\
J05055\,--5728 & 4763897739549071744 & 4763906879239461632 & -0.45  & [-0.49, -0.41]   & 0.80 & [0.64, 0.95] & 3760 \\
  $-$          & 4078432504018987904 & 4078432297860547072 & \;0.83 & [ 0.82, \, 0.84] & 1.53 & [1.29, 1.91] & 1584 \\
\hline

\end{tabular}
\label{tab:binary}

\end{table*}

\subsubsection{Expected encounter rates from stellar density}
To quantify the completeness of our encounter catalogue, we estimated the expected number of stellar encounters for different spectral types, accounting for their distinct kinematics. The expected number of encounters was calculated as the number of stars passing through a cylindrical interaction volume defined by a cross-sectional area $\pi d_\textrm{crit}^2$ and a length determined by the travel distance $tv_\textrm{S.T.}$, where $v_\textrm{S.T.}$ is the characteristic encounter velocity for a given spectral type and $t$ is the time interval considered. The expected number of encounters is given by $N=\sum_\textrm{S.T.}\pi d_\textrm{crit}^2\,tv_\textrm{S.T.}\,n_\textrm{S.T.}$ \citep{binney2011galactic}, where $n_\textrm{S.T.}$ is the number density of stars of a given spectral type. The characteristic encounter velocity is defined as $v_\textrm{S.T.}=\sqrt{v_{\beta\textrm{P,\;S.T.}}^2+\sigma_{v_\textrm{S.T.}}^2}$ \citep{torres2019}, where $v_{\beta\textrm{P,\;S.T.}}$ is the relative velocity of $\beta$ Pic with respect to the mean for stars of that spectral type, and $\sigma_{v_\textrm{S.T.}}$ is the corresponding velocity dispersion. We adopted Solar-relative velocities from \cite{1981GalaAstroMihalasBinney}, transformed into $\beta$ Pic'c reference frame, and used stellar densities and velocity dispersions from \cite{torres2019}. 

If we apply this method over the \new{3.5} Myr span between the earliest and the latest encounter, we find an expected total of \new{160} encounters\textemdash over five times the number actually identified (\new{49}). However, restricting the analysis to the better-sampled  $\pm$0.5 Myr interval, we find \new{30} encounters identified versus 46 expected, indicating a shortfall of \new{$\sim$35\%}. This incompleteness is consistent with the $\sim$38\% of stars in the GCNS sample lacking RVs, suggesting that our encounter recovery rate is broadly in line with the observational limitations.

\subsubsection{The 95\% probability threshold}
\label{subsub:pt95}
We conducted a self-consistency check to evaluate the impact of our 95\% confidence cut. Since we calculated the empirical probability for each star to come within 2 pc of $\beta$ Pic, summing these probabilities yields the expected number of such close passages over the whole statistical ensemble.

\begin{figure}[ht]

\centering
\includegraphics[width=0.45\textwidth]{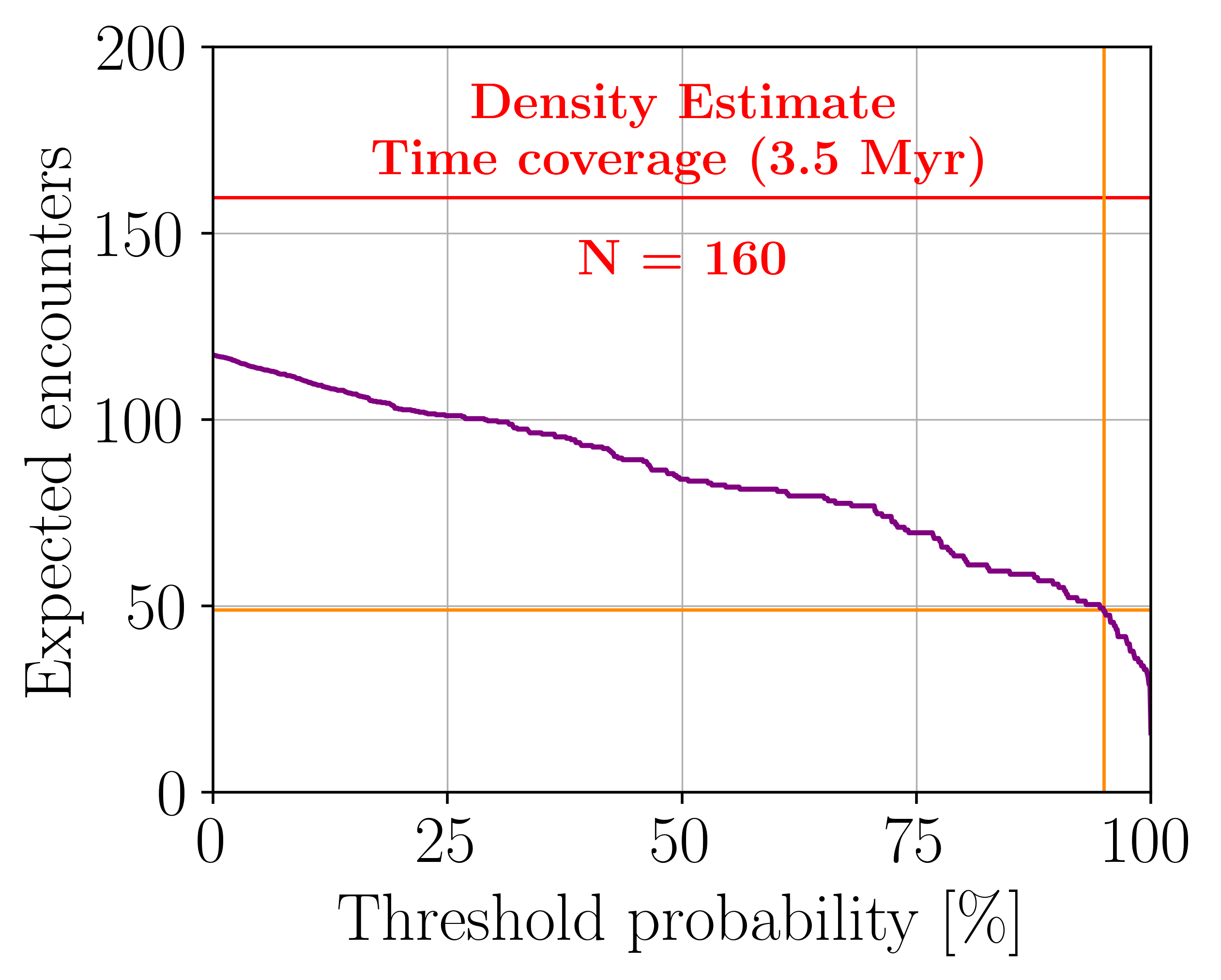}

\centering
\includegraphics[width=0.45\textwidth]{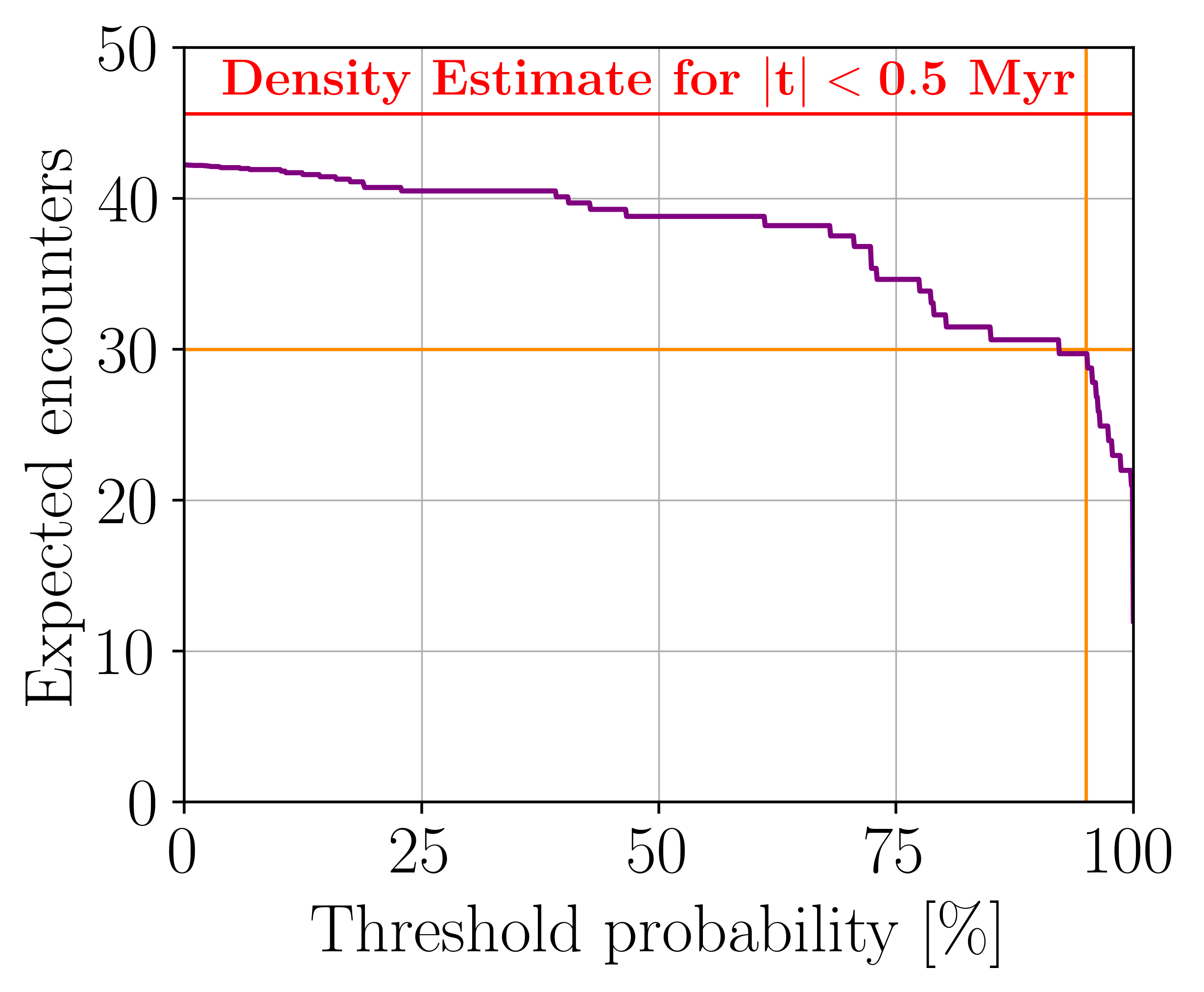}
\caption{Expected number of stars undergoing a close encounter (within 2 pc) with Beta Pictoris, as a function of the minimum encounter probability threshold. The upper panel corresponds to the entire time interval studied ($\sim$3.5 Myr), while the lower panel focuses on the most complete portion of the dataset, within the interval $\pm$0.5 Myr. The orange lines mark the 95\% probability threshold adopted in this work. At this threshold, the expected number of encounters is \renew{49} over 3.5 Myr and \renew{30} in the $\pm0.5\;\textrm{Myr}$ interval. The red line indicates the independent estimate obtained from the stellar density-based calculation, yielding 46 encounters in the $\pm$0.5 Myr interval. The total number of expected encounters from our data\textemdash corresponding to a $0\%$ probability threshold\textemdash is \renew{117} for the full sample and \renew{42} for the $\pm0.5\;\textrm{Myr}$ interval.
}
\label{fig:probdistt}
\end{figure}

As an example, four stars with probabilities of 2\%, 20\%, 80\%, and 98\% contribute a total of two expected encounters. Keeping all four with a low threshold (e.g. 1\%) increases completeness but risks false positives; a stricter threshold (e.g. 99\%) improves reliability but may exclude real events. The objective is to strike a balance between completeness and robustness, and we therefore adopt a 95\% threshold in this work.

By summing only those probabilities above various thresholds, we estimated how many real passages are expected at each confidence level. These results are shown in Figure \ref{fig:probdistt}, alongside the stellar-density-based estimates. We applied this analysis for both the full sample and the $\pm0.5\;\textrm{Myr}$ interval around the present, where the dataset is more complete.

Overall, our catalogue captures over \renew{$40\%$} of the expected close approaches across the full \renew{3.5} Myr, and \renew{more than $70\%$} within $\pm0.5\,\textrm{Myr}$. Future \textit{Gaia} data releases are expected to improve this by adding both new sources and better-quality RVs. Currently, RV uncertainties are, on average, ten times larger than those of tangential velocities derived from astrometry, indicating room for improvement in both RV quantity and quality.

\begin{table*}
\centering

\caption{\renew{Comparison of encounter parameters with those reported by \cite{kalas2001stellar}. The first two columns list the \textit{Hipparcos} IDs and the corresponding \textit{Gaia} DR3 IDs. To compare with \cite{kalas2001stellar} results, we computed time-distance uncertainties as the limits of the 2D $1 \sigma$ contours (39.35\%), based on Mahalanobis distances. Velocity uncertainties are given as $1 \sigma$ intervals of the marginal distributions (13.6$^\textrm{th}$-86.4$^\textrm{th}$ percentiles). The full table is available on \href{https://zenodo.org/records/15442718?preview=1&token=eyJhbGciOiJIUzUxMiJ9.eyJpZCI6IjNmODhjNTMwLTVmYjQtNDIzZC04MjZlLTk0MTVkN2Q2ZjkyMyIsImRhdGEiOnt9LCJyYW5kb20iOiJjMTU3OWE2ZjUxY2NiODY1OGIzZGNmZTdmNDZjMWUwOSJ9.157RkI0zdNQNui3Wn-vpk00-HvD1T5U7sUgQHVPn-_t5P5ZhYFeQicS0I_84AQatqVOEWiBXv_mL9A09lxcqZQ}{Zenodo}.}}

\begin{tabular}{cc cc cc cc }

\hline\hline
\noalign{\smallskip}

        \textit{Hipparcos} & \multirow{2}{*}{\textit{Gaia} DR3 ID} & 
        \multicolumn{2}{c}{Encounter Time [Myr]} & 
        \multicolumn{2}{c}{Encounter Distance [pc]} & 
        \multicolumn{2}{c}{Encounter Velocity [km s$^{-1}$]} \\ 
        \cmidrule(lr){3-4} \cmidrule(lr){5-6} \cmidrule(lr){7-8}
        ID & & Kalas et al. & This Work & Kalas et al. & This Work & Kalas et al. & This Work \\ 
        \hline

\hline
\noalign{\smallskip} 
10798  & 5117974602912370432\;\; & $ -0.318 _{-0.015} ^{+0.012}$ & $-0.346 _{-0.007} ^{+0.007}$ & $1.88  _{-0.40} ^{+0.51}$ & $1.27   _{-0.21}  ^{+0.22}$     & $42.8  \pm 1.3$     & $40.2 \pm 0.4$ \\
\noalign{\smallskip}  
17378  & 5164120762333028736\;\; & $ -0.295 _{-0.018} ^{+0.017}$ & $-0.320 _{-0.007} ^{+0.008}$ & $3.96  _{-0.41} ^{+0.42}$ & $3.94   _{-0.07}  ^{+0.07}$     & $47.9  \pm 1.5$     & $45.0 \pm 0.5$ \\
\noalign{\smallskip}  
19893  & 4781833626056162688\;\; & $ -0.031 _{-0.036} ^{+0.035}$ & $-0.024 _{-0.017} ^{+0.016}$ & $4.94  _{-0.14} ^{+0.19}$ & $5.02   _{-0.09}  ^{+0.09}$     & $13.4  \pm 1.4$     & $14.8 \pm 0.8$ \\
\noalign{\smallskip}  
19921  & 4678664766393829504\tablefootmark{1} & $ -0.119 _{-0.009} ^{+0.012}$ & $-0.101 _{-0.005} ^{+0.005}$ & $3.74  _{-0.33} ^{+0.31}$ & $4.19   _{-0.22}  ^{+0.21}$     & $28.6  \pm 0.6$     & $29.8 \pm 0.3$ \\
\noalign{\smallskip}  
22122  & 4784805056230587392\tablefootmark{2} & $ -0.034 _{-0.006} ^{+0.006}$ & $-0.034 _{-0.001} ^{+0.001}$ & $2.76  _{-0.18} ^{+0.56}$ & $2.74   _{-0.05}  ^{+0.05}$     & $62.7  \pm 1.5$     & $62.3 \pm 0.2$ \\
\noalign{\smallskip}  
23693  & 4763906879239461632\tablefootmark{1} & $ -0.356 _{-0.032} ^{+0.029}$ & $-0.447 _{-0.020} ^{+0.020}$ & $0.92  _{-0.12} ^{+0.13}$ & $0.78   _{-0.08}  ^{+0.08}$     & $21.6  \pm 1.5$     & $18.0 \pm 0.5$ \\
\noalign{\smallskip}  
25544  & 4759543295545939840\;\; & $ -0.116 _{-0.009} ^{+0.009}$ & $-0.123 _{-0.002} ^{+0.002}$ & $1.49  _{-0.30} ^{+0.39}$ & $1.26   _{-0.08}  ^{+0.08}$     & $25.5  \pm 0.9$     & $24.5 \pm 0.2$ \\
\noalign{\smallskip}  
27628  & 2887731882922767744\;\; & $ -0.107 _{-0.012} ^{+0.015}$ & $-0.108 _{-0.004} ^{+0.004}$ & $0.58  _{-0.11} ^{+0.51}$ & $0.75   _{-0.22}  ^{+0.21}$     & $83.9  \pm 2.1$     & $87.0 \pm 1.3$ \\
\noalign{\smallskip}  
29568  & 2913411183149615744\tablefootmark{1} & $ -0.693 _{-0.047} ^{+0.044}$ & $-0.743 _{-0.026} ^{+0.026}$ & $2.95  _{-0.32} ^{+0.90}$ & $4.93   _{-0.39}  ^{+0.38}$     & $11.6  \pm 0.5$     & $10.1 \pm 0.2$ \\
\noalign{\smallskip}  
29958  & 2993676867708999296\;\; & $ -0.198 _{-0.029} ^{+0.032}$ & $-0.177 _{-0.002} ^{+0.002}$ & $1.00  _{-0.40} ^{+2.36}$ & $3.27   _{-0.30}  ^{+0.31}$     & $102.1 \pm 5.6$\;\; & $97.4 \pm 0.8$ \\
\noalign{\smallskip}  
31711  & 5479778765278589568\tablefootmark{2} & $ -0.189 _{-0.076} ^{+0.050}$ & $-0.175 _{-0.046} ^{+0.047}$ & $3.97  _{-0.31} ^{+0.37}$ & $4.05   _{-0.13}  ^{+0.13}$     & $14.6  \pm 1.4$     & $17.5 \pm 1.2$ \\
\noalign{\smallskip}  
37504  & 5263150888430619904\;\; & $ -0.643 _{-0.038} ^{+0.053}$ & $-0.669 _{-0.022} ^{+0.022}$ & $4.59  _{-0.91} ^{+1.52}$ & $4.20   _{-0.52}  ^{+0.50}$     & $37.6  \pm 1.5$     & $40.5 \pm 0.4$ \\
\noalign{\smallskip}  
38908  & 5291028181119851776\tablefootmark{1} & $ -0.140 _{-0.009} ^{+0.006}$ & $-0.134 _{-0.002} ^{+0.002}$ & $1.97  _{-0.42} ^{+0.69}$ & $2.87   _{-0.24}  ^{+0.25}$     & $49.5  \pm 0.8$     & $48.1 \pm 0.2$ \\
\noalign{\smallskip}  
83990  & 5914096303621755520\;\; & $ -0.304 _{-0.015} ^{+0.018}$ & $-0.323 _{-0.005} ^{+0.005}$ & $3.92  _{-1.16} ^{+1.18}$ & $4.15   _{-0.22}  ^{+0.22}$     & $60.4  \pm 1.5$     & $57.0 \pm 0.4$ \\
\noalign{\smallskip}  
89042  & 6634032740544236032\tablefootmark{2} & $ -0.397 _{-0.023} ^{+0.020}$ & $-0.408 _{-0.003} ^{+0.003}$ & $2.29  _{-0.21} ^{+1.21}$ & $4.38   _{-0.31}  ^{+0.31}$     & $50.6  \pm 1.6$     & $48.3 \pm 0.3$ \\
\noalign{\smallskip}  
114996 & 6492406743706907776\tablefootmark{2} & $ -0.596 _{-0.043} ^{+0.033}$ & $-0.606 _{-0.056} ^{+0.057}$ & $3.94  _{-1.77} ^{+1.35}$ & $12.31  _{-1.86}  ^{+1.88}$\;\; & $29.3  \pm 0.8$     & $22.7 \pm 0.7$ \\
\noalign{\smallskip}  
116250 & 6353376831270492800\;\; & $ -0.884 _{-0.067} ^{+0.062}$ & $-0.853 _{-0.012} ^{+0.012}$ & $2.79  _{-1.92} ^{+2.39}$ & $1.38   _{-0.28}  ^{+0.27}$     & $22.9  \pm 1.7$     & $23.9 \pm 0.3$ \\
\noalign{\smallskip}  

\hline

\end{tabular}
\label{tab:kalas}
\tablefoot{\\
\tablefoottext{1}{According to SIMBAD, this star belongs to a binary or multiple system.}\\
\tablefoottext{2}{This \textit{Hipparcos}-\textit{Gaia} identification is not in the \textit{Gaia} DPAC crossmatch tables.}\\
}
\end{table*}

\section{Comparison with \texorpdfstring{\cite{kalas2001stellar}}{Kalas et al. (2001)}}
\label{sec:kalas}
The first systematic search for stellar encounters with the $\beta$ Pic system was carried out by \cite{kalas2001stellar}, who used astrometric data from the \textit{Hipparcos} catalogue \citep{perryman1997hipparcos} and RVs from \cite{barbier2000catalogue}. By propagating linear trajectories for a sample of $21\,497$ stars, they identified 18 candidates that passed within 5 pc of $\beta$ Pic over the past 1 Myr. In contrast, our study using \textit{Gaia} data reveals \renew{146} in the same time span and distance range\textemdash nearly an order of magnitude more. This increase reflects both the superior precision and improved completeness of the \textit{Gaia} dataset. Nonetheless, based on stellar density estimates, that sample is only 50\% complete (see Figure \ref{fig:probdistt}).

To directly compare our results with those of \cite{kalas2001stellar}, \renew{we computed the 1$\sigma$ confidence intervals for the time and distance of closest approach using
the limits that enclose the $39.35\%$ clones with the lower Mahalanobis distance to the distribution.} For the relative velocities at closest approach, we used the 68.27\% confidence interval, defined by the $15.87^\textrm{th}$ and the $84.14^\textrm{th}$ percentiles of the marginal velocity distributions. These values are reported in Table \ref{tab:kalas} for the 17 stars from \cite{kalas2001stellar} for which \textit{Gaia} astrometric solutions are available.

One star from the original sample, HIP 93506 (GDR3 6760703042771435136), is absent from our catalogue due to the lack of \textit{Gaia} astrometry. Additionally, HIP 22122, HIP 31711, HIP 89042, and HIP 114996 are not present in the \textit{Hipparcos–Gaia} crossmatch tables, likely due to low-quality \textit{Hipparcos} astrometric solutions that prevented robust identification in \textit{Gaia}. For these objects, we adopted the SIMBAD identification.

Among the \renew{49} encounters listed in Table \ref{tab:encounters}, only 5 overlap with those from \cite{kalas2001stellar}. It is important to note that \cite{kalas2001stellar} used a cutoff of 5 pc. They reported 6 stars that have a passage closer than 2 pc with $\beta$ Pic. Applying our 95\% confidence criterion excludes HIP 114996 (GDR3 6492406743706907776), while \cite{kalas2001stellar} also identified HIP 29958 (GDR3 2993676867708999296) and HIP 38908 (GDR3 5291028181119851776) as having close encounters within 2 pc\textemdash cases not supported by our results.

A search in SIMBAD reveals that the \cite{kalas2001stellar} candidates HIP 19921, HIP 23693, HIP 29568, and HIP 38908 belong to multiple systems. Unresolved multiplicity can distort the astrometry reported in the \textit{Hipparcos} catalogue, likely contributing to inaccuracies in the derived stellar trajectories.

\section{Summary and conclusions}
\label{sec:concl}
We have compiled the most comprehensive catalogue to date of stellar encounters with the $\beta$ Pictoris system by reconstructing the orbits of nearly $100\,000$ nearby stars using \textit{Gaia} DR3 astrometry and radial velocity data from \textit{Gaia} and other major spectroscopic surveys. From this dataset, we identified 49 stars (Table \ref{tab:encounters}) that either have passed or will pass within 2 pc of $\beta$ Pic, each with a probability greater than $95\%$, based on Monte Carlo orbital propagations. Based on stellar density estimates, we assess our sample to be about \renew{65\%} complete within the time interval $\pm$0.5 Myr. Notably, despite $\beta$ Pic being the eponym of its famous moving group, none of these encounters involved $\beta\,$PMG members; all were with unrelated field stars. 

\new{Among the \new{49} encounters, we identified \new{41} single stars and \new{8 members of five binary systems.} Four of these systems (WDS J03572-4413,
WDS J13237+0243, WDS J05055-5728 and the pair GDR3 4078432504018987904 / GDR3 4078432297860547072) \new{show coherent centre-of-mass trajectories,} suggesting a genuine close encounter with $\beta$ Pic. Multiplicity in stellar perturbers can enhance their gravitational influence, highlighting the need for sophisticated dynamical treatments.}

\renew{Although GDR3 1311454726097258368 passes closest to $\beta$ Pictoris (0.442 pc), its relatively low mass (0.07 $M_\odot$) and high velocity (62.16 km s$^{-1}$) make its perturbation negligible, insufficient to shape the $\beta$ Pictoris disc or account for its observed features. The stellar encounters we identify are therefore expected to affect only the distant outskirts of the system, while the disc structures revealed by JWST imaging \citep{rebollido2024jwst} are more likely the result of internal dynamical processes \cite[e.g.,][]{beust2024dynamics}.}


This work is limited by the initial data used, particularly in the radial velocities. One third of our sample near $\beta$ Pic lacks RVs, and RV errors are typically ten times larger than those in tangential velocities, derived from proper motions and parallaxes. We expect future \textit{Gaia} data releases to increase the number of stars with available RVs. While planet-hunting radial velocity surveys offer higher accuracy, they are only available for a small subset of stars compared to the \textit{Gaia} catalogues, and the heterogeneity of sources complicates their systematic use.

Future work should incorporate detailed N-body simulations, including binary perturbers, and high particle counts that lead to strong statistical conclusions. The encounter catalogue presented here provides a robust foundation for future dynamical studies of the $\beta$ Pictoris system and, more generally, of how stellar encounters shape the architecture of planetary systems.

\begin{acknowledgements}

We thank the referee for their suggestions and comments, which helped us improve the quality and clarity of the paper. JLGM and EV acknowledges the support from the Spanish Ministry of Science and Innovation/State Agency of Research (MCIN/AEI) under the grant PID2021-127289-NB-I00. ST acknowledges the funding from the European Union’s Horizon 2020 research and innovation program under the Marie Sk\l{}odowska-Curie grant agreement No 101034413. AJM acknowledges support from the Swedish National Space Agency (Career Grant 2023-00146) and from the Swedish Research Council (Project Grant 2022-04043).

This work has made use of data from the European Space Agency (ESA) mission \textit{Gaia} (https://www.cosmos.esa.int/gaia), processed by the \textit{Gaia} Data Processing and Analysis Consortium (\href{https://www.cosmos.esa.int/web/gaia/dpac/consortium}{DPAC}). Funding for the DPAC has been provided by national institutions, in particular the institutions participating in the \textit{Gaia} Multilateral Agreement.

This research has made use of the SIMBAD database,
operated at CDS, Strasbourg, France.

This project makes extensive use of the \href{https://numpy.org/}{NumPy} library for numerical computations. We are grateful to the developers and maintainers of NumPy for their invaluable work.

This work made use of \href{http://www.astropy.org}{Astropy}, a community-developed core Python package and an ecosystem of tools and resources for astronomy.

\end{acknowledgements}

%
\bibliographystyle{aa} 
\bibliography{refs} 
\begin{appendix}




\onecolumn

\section{TOPCAT Queries}
\label{ap:topcat}
To select the sphere around $\beta$ Pic, we used the heliocentric positions of each star, $\mathbf{r_\star}$, and $\beta$ Pic, $\mathbf{r_{\beta\mathrm{P}}}$, to checked whether the distance is less than $80 \,\mathrm{pc}$, ($(\mathbf{r_\star} - \mathbf{r_{\beta\mathrm{P}}})^2 < (80 \,\mathrm{pc})^2$):
\begin{verbatim}
dotProduct(
  subtract(astromXYZ(RA_ICRS, DE_ICRS, Plx),
           astromXYZ(86.8212345201, -51.0661362578, 50.9307)),
  subtract(astromXYZ(RA_ICRS, DE_ICRS, Plx),
           astromXYZ(86.8212345201, -51.0661362578, 50.9307))
) < 80*80
\end{verbatim}
Here, the function \texttt{astromXYZ} calculates the Cartesian components of the position in parsecs from right ascension and declination (in degrees) and parallax (in milliarcseconds).\\

To select GDR3 radial velocity data over GCNS, we used the error value as an indicator of data availability:
\begin{verbatim}
RV_o   = eRV_dr3 > 0 ? RV_dr3  : RV_GCNS         # Radial velocity
eRV_o  = eRV_dr3 > 0 ? eRV_dr3 : eRV_GCNS        # Radial velocity uncertainty
r_RV_o = eRV_dr3 > 0 ? "GDR3"  : r_RV_GCNS       # Radial velocity reference
\end{verbatim}

\begin{figure}[ht]

\centering
\includegraphics[width=0.7\textwidth]{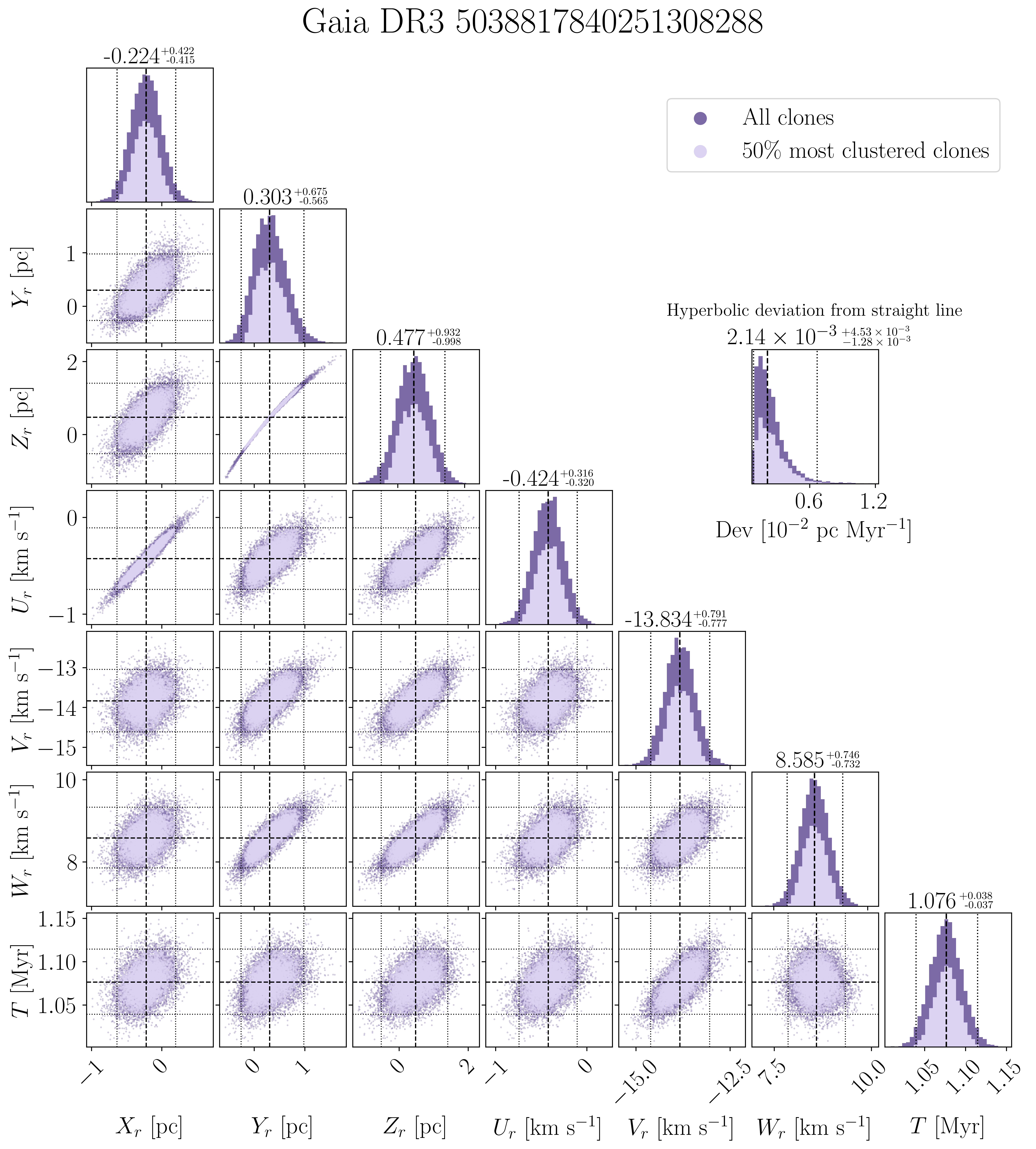}
\caption{\renew{Encounter parameters ($X,Y,Z,U,V,W,t$) of the clones of GDR3 5038817840251308288 relative to $\beta$ Pictoris. The 50\% most clustered clones, selected using the Mahalanobis distance, are highlighted. The values above the histograms indicate the median ($x^{50}$) and the 95\% confidence interval $[x^{2.5},x^{97.5}]$.}}
\label{fig:corner}
\end{figure}
\newpage
To select the data from the survey with the smallest error:
\begin{verbatim}
RV_o   = (eRV_i < eRV_a) | !(eRV_a > 0) ? RV_i   : RV_a
eRV_o  = (eRV_i < eRV_a) | !(eRV_a > 0) ? eRV_i  : eRV_a
r_RV_o = (eRV_i < eRV_a) | !(eRV_a > 0) ? r_RV_i : "ALTERNATIVE_SURVEY_NAME"
\end{verbatim}
Here, the subscript ``\_i'' refers to the input survey, ``\_a'' to the alternative survey whose error is being compared, and ``\_o'' to the resulting parameters.

\section{Clone set example}
\label{ap:clones}
\renew{This appendix presents an example outcome of the orbital integrations: the clone distribution of relative positions, velocities, and encounter times for GDR3 5038817840251308288 with $\beta$ Pictoris. The corresponding corner plot is shown in Figure \ref{fig:corner}. Corner plots for the remaining high-confidence encounters, as well as for the centres of mass of the binary systems, are available at \href{https://zenodo.org/records/17177886?preview=1&token=eyJhbGciOiJIUzUxMiJ9.eyJpZCI6IjQ0N2Y0MjdhLTVhNWQtNGIwZC04YWE5LWU3OWU3NGJkNmMzZSIsImRhdGEiOnt9LCJyYW5kb20iOiJlYTA5ZGE4YWRiNjE1NGVkYjZjMjg4YzU0MDg5ZTcwYiJ9.MdZxvxm5K1lp-G_qTQE-1uinkYpMAUVDumVeYZbVPfm_2L4pMuSj2fyG0MZEM5Z96DQ1CRm21-5FALU7jRBf9g}{Zenodo}.  }

\end{appendix}
\end{document}